\documentclass[letterpaper]{article}

\usepackage[utf8]{inputenc}
\usepackage{arxiv}
\usepackage{float}
\usepackage{enumitem}
\usepackage[english]{babel}
\usepackage{amssymb}
\usepackage{amsmath,mathtools}
\usepackage{stmaryrd} 								
\usepackage[table]{xcolor}
\usepackage{graphicx}
\usepackage{multirow}
\usepackage{cprotect}
\usepackage{bbold}
\usepackage{overpic}
\usepackage{psfrag}
\usepackage{bm}
\usepackage{lineno}
\usepackage{url}
\usepackage{physics}								
\usepackage{epsfig}
\usepackage{marvosym}
\usepackage{verbatim}
\usepackage{subcaption}
\usepackage{caption}
\usepackage{hyperref}
\hypersetup{colorlinks=true,linkcolor=red,citecolor=blue}
\usepackage{ulem}
\usepackage{lmodern}
\usepackage{booktabs} 
\usepackage{moresize} 
\usepackage{siunitx}
\usepackage{mathtools}
\usepackage{tikz}
\usepackage{pgfplots}
\pgfplotsset{compat=1.14}
\usepgfplotslibrary{fillbetween}
\usetikzlibrary{shapes, arrows, arrows.meta, calc, positioning, decorations,
decorations.pathmorphing, decorations.text, spy}
\pgfplotsset{translate gnuplot=true}
\usepackage{etex} 								
\usepgfplotslibrary{external} 						
\tikzsetexternalprefix{ext/}							
\usepackage{booktabs}
\usepackage{makecell}
\usepackage{rotating,tabularx}

\definecolor{blue1}		{RGB}{0,177,234}				
\definecolor{blue2}		{RGB}{76,200,239}				
\definecolor{blue3}		{RGB}{127,215,244}				
\definecolor{blue4}		{RGB}{178,231,248}				
\definecolor{bluegray1}	{RGB}{0,127,167}				
\definecolor{bluegray2}	{RGB}{76,165,193}				
\definecolor{bluegray3}	{RGB}{127,191,211}				
\definecolor{bluegray4}	{RGB}{178,216,228}				
\definecolor{gray1}		{RGB}{76,84,93}				
\definecolor{gray2}		{RGB}{129,135,141}				
\definecolor{gray3}		{RGB}{165,169,174}				
\definecolor{gray4}		{RGB}{201,203,206}				
\definecolor{gray5}		{RGB}{230,230,230}				
\definecolor{gray6}		{RGB}{245,245,245}				
\definecolor{orange1}	{RGB}{255,126,46}				
\definecolor{orange2}	{RGB}{255,164,108}				
\definecolor{orange3}	{RGB}{255,190,150}				
\definecolor{orange4}	{RGB}{255,216,192}				
\definecolor{brown1}		{RGB}{205,133,63}				
\definecolor{green1}		{RGB}{0,168,107}				
\definecolor{green2}		{RGB}{30,198,137}				
\definecolor{green3}		{RGB}{60,228,167}				
\definecolor{green4}		{RGB}{90,255,197}				
\definecolor{purple1}		{RGB}{89,89,171}				
\definecolor{purple2}		{RGB}{159,159,195}				
\definecolor{purple3}		{RGB}{189,189,225}				
\definecolor{purple4}		{RGB}{209,209,255}				
\definecolor{teal1}		{RGB}{0,128,128}				
\definecolor{teal2}		{RGB}{100,168,168}				
\definecolor{teal3}		{RGB}{130,198,198}				
\definecolor{teal4}		{RGB}{160,228,228}				
\definecolor{red1}		{RGB}{255,0,0}					
\definecolor{red2}		{RGB}{255,30,30}				
\definecolor{red3}		{RGB}{255,60,60}				
\definecolor{red4}		{RGB}{255,90,90}				
\definecolor{royal1}		{RGB}{2,119,189}				
\definecolor{royal2}		{RGB}{32,149,219}				
\definecolor{royal3}		{RGB}{62,179,249}				
\definecolor{royal4}		{RGB}{92,209,255}				

\pgfplotsset{
    colormap={custom_map}{[5pt]
            rgb255(0pt)=(255,126,46);
            rgb255(500pt)=(255,190,150);
            rgb255(1000pt)=(0,177,234);
            rgb255(1500pt)=(127,215,244);
    },
}

\newcommand\ie{\textit{i.e.}\,\,}

\renewcommand{\emph}[1]{\textit{#1}}
\setenumerate{label=\textcolor{bluegray1}{$\bm{\diamond}$},itemsep=0.0pt,topsep=5pt}

\newcommand{\eps}{\varepsilon}
\newcommand{\V}[1]{\bm{#1}}
\newcommand{\pr}{\text{Pr}}									
\newcommand{\ra}{\text{Ra}}									
\newcommand{\pe}{\text{Pe}}									
\newcommand{\re}{\text{Re}}									
\newcommand{\nus}{\text{Nu}}									
\newcommand{\iu}{\mathrm{i}}									

\title{\textsc{beacon}, a lightweight deep reinforcement learning benchmark library for flow control}

\def\size{7.3cm}
\author{
	\parbox{\size}{\centering J. Viquerat\thanks{Corresponding author}}\\
	MINES Paristech, CEMEF\\
	PSL - Research University\\
	\texttt{jonathan.viquerat@mines-paristech.fr}\\
\And
	\parbox{\size}{\centering P. Meliga}\\
	MINES Paristech, CEMEF\\
	PSL - Research University
\And
	\parbox{\size}{\centering P. Jeken-Rico}\\
	MINES Paristech, CEMEF\\
	PSL - Research University
\And
	\parbox{\size}{\centering E. Hachem}\\
	MINES Paristech, CEMEF\\
	PSL - Research University
}

\begin{document}
\newgeometry{left=3cm,right=3cm,top=3cm,bottom=2.5cm}
\maketitle

\begin{abstract}
Recently, the increasing use of deep reinforcement learning for flow control problems has led to a new area of research, focused on the coupling and the adaptation of the existing algorithms to the control of numerical fluid dynamics environments. Although still in its infancy, the field has seen multiple successes in a short time span, and its fast development pace can certainly be partly imparted to the open-source effort that drives the expansion of the community. Yet, this emerging domain still misses a common ground to (i) ensure the reproducibility of the results, and (ii) offer a proper \textit{ad-hoc} benchmarking basis. To this end, we propose \textsc{beacon}, an open-source benchmark library composed of seven lightweight \textcolor{black}{one-dimensional and two-dimensional} flow control problems with various characteristics, action and observation space characteristics, and CPU requirements. In this contribution, the seven consideblack problems are described, and reference control solutions are provided. The sources for the following work are available at \url{https://github.com/jviquerat/beacon}.
\end{abstract}


\section{Introduction}

In recent years, the area of deep reinforcement learning-based flow control has undergone a rapid development, with a surge of contributions on topics such as (but not limited to) drag blackuction~\cite{yzwang2022}, collective swimming~\cite{novati2017} or heat transfers~\cite{beintema2020}. \textcolor{black}{Unlike traditional methods, deep reinforcement learning (DRL) enables the learning of complex control strategies directly from data, thereby alleviating the effects of local minima and generalizability of algorithm towards other scenarios~\cite{viquerat2021}.} Yet, the inherent reproducibility issues of \textcolor{black}{DRL} algorithms~\cite{andrychowicz2020}, as well as the variety of \textcolor{black}{computational fluid dynamics (CFD)} solvers and the possible variability of environment design among the different actors of the community make it hard to accurately compare algorithm performances, thus hindering the general progress of the field. More, the standard DRL benchmarks (such as the \textsc{mujoco} package~\cite{mujoco}, or the Atari games from the \textcolor{black}{arcade learning environments (\textsc{ale})}~\cite{atari}) have a limited interest in the context of benchmarking DRL methods for flow control, as their dynamics, observation spaces, computational requirements and action constraints display substantial differences with those of numerical flow control environments. 

\textcolor{black}{From a general point of view, flow control problems are characterized by a simulated physics environment spanned over at least two dimensions, possibly including time. The control is performed by an agent that modifies boundary conditions, source terms or other components of the domain in order to optimize a given objective. A notable difficulty is hereby designing a robust, and at the same time efficient environment that is able to cope with a wide range of actions while preserving low and stable runtimes~\cite{gym}. One way of minimizing the computational cost is lumping the Navier-Stokes equations, the backbone of most fluid mechanics problems, by limiting the dimensionality and restricting its terms to the dominant ones for each problem at hand. This way, it is possible to retain the main features of the flow, while tuning the schemes and discretizations towards higher performances.}

In the present contribution, we lay the first stone of a numerical flow control benchmark library for DRL algorithms to systematically assess methodological improvements on physically and numerically relevant problems. The design of the test cases voluntarily limits the computational cost of the solvers, making this library a first benchmarking step before testing on more complex and CPU-intensive cases.

The organization is as follows: a short presentation of the library and its general characteristics is proposed in section~\ref{section:lib}, after what the environments are introduced in a systematic way in section~\ref{section:envs}. For each case, the physics of the problem are described, followed by insights on the discretization. Then, the environment parameters and specificities are described, after what baseline learning curves and details on the solved environment are provided. Finally, the perspectives for the present work are exposed in a conclusive section.

\section{The \textsc{beacon} library}
\label{section:lib}

This library provides self-contained cases for deep reinforcement learning-based flow control. The goal is to provide the community with benchmarks that fall within the range of flow control problems, while following three constraints: (i) be written in Python to ensure a simple coupling with most DRL implementations, (ii) follow the general \textsc{gym} \textcolor{black}{application programming interface~\cite{gym}}, and (iii) be cheap enough in terms in CPU usage so that trainings can be performed on a decent computing station. \textcolor{black}{Aligned with the standardized approach of \textsc{gym}, which streamlines environment setup and facilitates a focused exploration of RL research, this library serves as a first step for prototyping flow control algorithms before moving on to larger problems that will require more efficient CFD solvers and, most probably, a CPU cluster.}

The original version of the library contains seven cases, whose main characteristics are presented in table~\ref{table:environments}. The selection of the cases was made in order to (i) follow the aforementioned contraints and (ii) propose a variety of problem (episodic or continuous) and control (discrete or continuous) types, as well as different action space dimensionalities. For each case, some parameters can be tuned that can significantly modify the difficulty and the CPU requirements of the problem. In their default configurations, two cases have low CPU requirements and can be run on a standard laptop; three have intermediate computational loads and will require an extended running time on a laptop or a workstation, and can benefit from the use of parallel environments; finally, two have high computational needs and will require a decent workstation and parallel sample collection~\cite{rabaultkuhnle2019, viquerat2023}. Some cases, such as \verb!rayleigh-v0!, \verb!lorenz-v0!~\cite{beintema2020}, \verb!vortex-v0!~\cite{meliga2011b} and \verb!shkadov-v0!~\cite{belus2019}, were taken from the literature and fully re-implemented, while others were designed specifically for this work. All the environments of the library follow similar development pattern, and are self-contained to simplify re-usability. Provided that the Python language is not optimal in terms of performance, the core routines are deferblack to \textsc{numba}~\cite{numba} to blackuce the execution time. To avoid version conflicts and improve compatibility, additional package requirements are strictly blackuced to \textsc{gym}~\cite{gym}, \textsc{numpy}~\cite{numpy} and \textsc{matplotlib}~\cite{matplotlib}.

\begin{table}[h]
    \footnotesize
    \caption{\textbf{Overview of the different environments} in their default configurations.}
    \label{table:environments}
    \centering
    \begin{tabular}{rcccc}
        \toprule
        env. name			& action dim.	& CPU requirements 	& problem type		& control type	\\\midrule
	\verb!shkadov-v0!	& $5$		& moderate			& continuous		& continuous	\\
	\verb!rayleigh-v0!	& $10$		& high				& continuous		& continuous 	\\
	\verb!mixing-v0!	& $4$		& high				& episodic			& discrete		\\
	\verb!lorenz-v0!		& $1$		& moderate			& continuous		& discrete		\\
	\verb!burgers-v0!	& $1$		& low				& continuous		& continuous	\\
	\verb!sloshing-v0!	& $1$		& low				& episodic			& continuous	\\
	\verb!vortex-v0!		& $2$		& moderate			& continuous		& continuous	\\
        \bottomrule
    \end{tabular}
\end{table}

All environments are solved with an in-house implementation of the \textcolor{black}{proximal policy optimization (\textsc{ppo})} algorithm~\cite{ppo}, for which the default parameters are provided in table~\ref{table:default_ppo_parameters}. Depending on the control type, results obtained from an off-policy algorithm (either \textcolor{black}{deep Q-networks (\textsc{dqn}) \cite{dqn} or time-delayed deep deterministic policy gradient (\textsc{td3}) \cite{td3}}) are also shown for comparison. The performances of these algorithms are evaluated on standard benchmarks in appendix \ref{section:benchmark}.

\begin{table}[h]
    \footnotesize
    \caption{\textbf{Default parameters used for the} \textsc{ppo} \textbf{agent.}}
    \label{table:default_ppo_parameters}
    \centering
    \begin{tabular}{rll}
        \toprule
        --					& agent type						& \textsc{ppo}-clip\\
	$\gamma$ 			& discount factor					& 0.99\\
	$\lambda_a$ 			& actor learning rate					& \num{5e-4}\\
	$\lambda_c$ 			& critic learning rate					& \num{5e-3}\\
	--		 			& optimizer						& adam\\
	--					& weights initialization				& orthogonal\\
	--	 				& activation (actor hidden layers)		& tanh\\
	-- 					& activation (actor final layer, continous)	& tanh, sigmoid\\
	-- 					& activation (actor final layer, discrete)	& softmax\\
	--	 				& activation (critic hidden layers)		& relu\\
	-- 					& activation (critic final layer)			& linear\\
	$\epsilon$ 			& PPO clip value					& 0.2\\
	$\beta$				& entropy bonus					& 0.01\\
	$g$					& gradient clipping value				& 0.1\\	
	-- 					& actor network						& $[64, [[64],[64]]]$\\
	-- 					& critic network						& $[64, 64]$\\
	--					& observation normalization			& yes\\
	--					& observation clipping				& no\\
	--					& advantage type					& GAE\\
	$\lambda_\text{GAE}$	& bias-variance trade-off				& 0.99\\
	--					& advantage normalization			& yes\\\midrule
	$n_\text{rollout}$ 		& nb. of transitions per update			& env-specific\\
	$n_\text{batch}$ 		& nb. of minibatches per update 		& env-specific\\
	$n_\text{epoch}$		& nb. of epochs per update			& env-specific\\
	$n_\text{max}$			& total nb. of transitions per training		& env-specific\\
	$n_\text{training}$		& total nb. of averaged trainings		& 5\\
        \bottomrule
    \end{tabular}
\end{table}

\clearpage

\section{Environments}
\label{section:envs}

\subsection{Shkadov}

\subsubsection{Physics}

\textcolor{black}{The initial investigation into vertically falling fluid films was conducted by Kapitza \& Kapitza~\cite{kapitza1948}, sparking extensive experimental exploration in subsequent decades. These experiments revealed that waves on the surface of a descending thin liquid film exhibit strong non-linearity, manifesting the emergence of saturated waves from small perturbations in amplitude, as well as the presence of solitary waves. For low Reynolds numbers ($Re < 300$), it was noted that the wavelength of the non-linear waves greatly exceeded the thickness of the film, allowing for potential simplifications in their physical modeling (referblack to as the long-wave regime). Various physical models were proposed, among them the Shkadov model, which was introduced in 1967~\cite{shkadov1967}. Despite being shown to be inconsistent~\cite{lavalle2014}, the model exhibits intriguing spatio-temporal dynamics while remaining computationally affordable. The flow rate $q$ and the fluid height $h$ are simultaneously evolved using the following set of equations:}

\begin{equation}
\label{eq:shkadov}
\begin{split}
	\partial_t h 	&= -\partial_x q, \\
	\partial_t q		&= - \frac{6}{5} \partial_x \left( \frac{q^2}{h} \right) + \frac{1}{5 \delta} \left(  h \left( 1 + \partial_{xxx}h \right) - \frac{q}{h^2} \right),
\end{split}
\end{equation}

\begin{figure}
\centering
\pgfdeclarelayer{background}
\pgfsetlayers{background,main}
\begin{tikzpicture}[	scale=0.8, trim axis left, trim axis right, font=\scriptsize]
	\begin{axis}[	xmin=0, xmax=500, ymin=0.5, ymax=3.5, scale=1.0,
				width=\textwidth, height=.25\textwidth, scale only axis=true,
				legend cell align=left, legend pos=north east,
				grid=major, xlabel=$x$, ylabel=$h$]

		\begin{pgfonlayer}{background}
			\fill[color=bluegray2,opacity=0.3] (axis cs:0,0.5) rectangle (axis cs:150,5);
			\fill[color=orange2,opacity=0.3] (axis cs:150,0.5) rectangle (axis cs:275,5);
			\fill[color=teal2,opacity=0.3] (axis cs:275,0.5) rectangle (axis cs:500,5);
		\end{pgfonlayer}
		
		\addplot[draw=gray1, very thick, smooth] 			table[x index=0,y index=1] {shkadov_free.dat};
		\addplot[draw=black, thick, dash pattern=on 2pt] 	coordinates {(150,0.5) (150,5)}; 
		\addplot[draw=black, thick, dash pattern=on 2pt] 	coordinates {(275,0.5) (275,5)};
			
	\end{axis}
\end{tikzpicture}
\caption{\textbf{Example of developed flow for the Shkadov equations with $\delta = 0.1$.} Three regions can be identified: a first region where the instability grows from a white noise (blue), a second region with pseudo-periodic waves (orange), and a third region with non-periodic, pulse-like waves (green).} 
\label{fig:shkadov_free}
\end{figure}
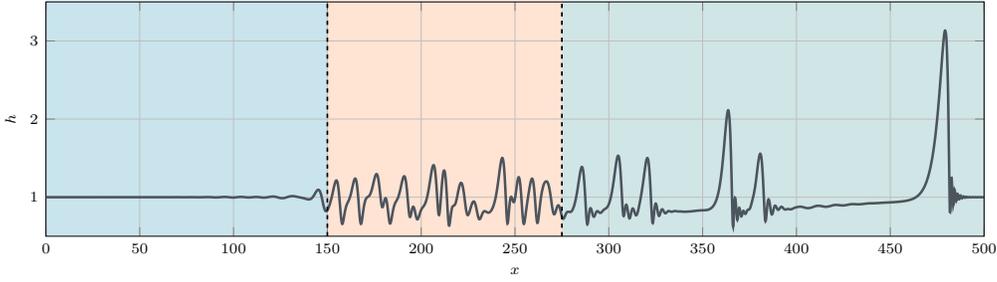

\noindent \textcolor{black}{where the $\delta$ parameter encompasses all the physics of the problem:}

\begin{equation}
\label{eq:shkadov_delta}
	\delta = \frac{1}{15} \left( \frac{3 {Re}^2}{W} \right)^{\frac{1}{3}},
\end{equation}

\noindent with $Re$ and $W$ are the Reynolds and the Webber numbers, respectively defined on the flat-film thickness and the flat-film average velocity~\cite{chang2002}. The system (\ref{eq:shkadov}) is solved on a 1D domain of length $L$, with the following initial and boundary conditions:

\begin{equation}
\label{eq:shkadov_bc}
\begin{split}
	q(x,0)		= 1 &\text{ and } h(x,0)	= 1, \\
	q(0,t) 		= 1 &\text{ and } h(0,t) 	= 1 + \mathcal{U} ( -\eps, \eps), \\
	\partial_x q(L,t) = 0 &\text{ and } 	\partial_x q(L,t) = 0,
\end{split}
\end{equation}

with $\eps \ll 1$ being the noise level \textcolor{black}{generated by the uniform distribution $\mathcal{U}$}. \textcolor{black}{As depicted in figure~\ref{fig:shkadov_free}, the introduction of random uniform noise at the inlet initiates the onset of exponentially growing instabilities (blue region), which subsequently exhibit a pseudo-periodic pattern (orange region). In the subsequent area, the periodicity of the waves breaks, and the instabilities transform into pulse-like formations, characterized by a sharp leading edge followed by minor ripples~\cite{chang2002book}. Certain of these sharp pulses, known as solitary pulses, move faster than others, leading to the amalgamation of upstream pulses in coalescence events. The parameter $\delta$ completely governs the dynamics of these solitary pulses, while the position of the transition region may also be linked to the level of noise at the inlet~\cite{chang2002}. It's worth noting that the Shkadov equations bear a close relationship with the Kuramoto-Sivashinsky system, which was independently discoveblack later by Kuramoto and Sivashinsky~\cite{koulago1995}.}

\subsubsection{Discretization}

\textcolor{black}{Equations (\ref{eq:shkadov}) undergo discretization employing a finite difference methodology. Given the presence of abrupt gradients, the convective terms are discretized using a Total Variation Diminishing (TVD) scheme incorporating a minmod flux limiter. The third-order derivative approximation is achieved by combining a second-order centeblack difference for the second derivative with a second-order forward difference, resulting in the following second-order approximation:}

\begin{equation}
\label{eq:shkadov_fd}
\begin{split}
	\left. \partial_{xxx} h \right|_j 	&\sim \frac{1}{2 \Delta x^3} \left( -h_{j+3} + 6h_{j+2} - 12h_{j+1} + 10 h_j - 3 h_{j-1} \right) \\
							&= \partial_{xxx} h (x_j) + \frac{\Delta x^2}{4} \partial_{xxxxx} h (x_j) + O\left( \Delta x^3 \right).
\end{split}
\end{equation}

\textcolor{black}{A second-order Adams-Bashforth method is used to integrate the system in time}:

\begin{equation}
\label{eq:adams_bashforth}
	h^{n+1} = h^n + \frac{\Delta t}{2} \left( - 3 f_h \left(h^{n} \right) + f_h \left(h^{n-1} \right) \right),
\end{equation}

\noindent where $f_h$ represents the right hand side of the evolution equation of $h$ in system (\ref{eq:shkadov}), and similarly for the $q$ field.

\subsubsection{Environment}

\textcolor{black}{The provided setup is a re-implementation based on the original work by Belus \textit{et al.}~\cite{belus2019}, albeit with several distinctions. Notably, the translational invariance aspect introduced in \cite{belus2019} is \emph{not} utilized in this context. Instead, this control scenario is viewed as an opportunity to evaluate algorithms dealing with problems featuring arbitrarily high action dimensionality.}

\textcolor{black}{For the control of the system (\ref{eq:shkadov}), a forcing term $\delta q_j$ is introduced into the equation governing the temporal evolution of the flow rate. Practically, localized jets are inserted at specific locations within the domain, as illustrated in figure~\ref{fig:shkadov_fields} below. The intensities of these jets are determined by the DRL agent. Initially, the first jet is positioned at $x_0=150$, with the default jet spacing set to $\Delta x_\text{jets} = 10$, akin to~\cite{belus2019}. To conserve computational resources, the domain length is conditioned by the number of jets $n_\text{jets}$ and their spacing:}

\begin{equation}
	L = L_0 + \left( n_\text{jets} + 2\right) \Delta x_\text{jets}.
\end{equation}

\textcolor{black}{By default, $L_0$ is set equal to $150$ (which corresponds to the beginning of the pseudo-periodic region for $\delta = 0.1$), and $n_\text{jets}$ is initialized to 1. The spatial discretization step is designated as $\Delta x = 0.5$, while the numerical time step is $\Delta t = 0.005$ time units. The inlet noise level is configublack as $\eps = \num{5e-4}$, mirroring the setup in~\cite{belus2019}. The injected flow rate $\delta q_j$ takes the following form:}

\begin{equation}
\label{eq:shkadov_jets}
	\delta q_j (x,t) = A u_j(t) \frac{4 \left( x - x_j^l \right) \left( x_j^r - x \right)}{ \left( x_j^r - x_j^l \right)^2},
\end{equation}

\textcolor{black}{with $A=5$ an \textit{ad-hoc} non-dimensional amplitude factor, $x_j^l$ and $x_j^r$ representing the left and right limits of jet $j$, and $u_j(t) \in [-1,1]$ denoting the agent-provided jet amplitude. Equation (\ref{eq:shkadov_jets}) corresponds to a parabolic profile of the jet in $x$, ensuring a zero flow rate at the boundaries. The jet width $x_j^r - x_j^l$ is fixed at $4$, in line with~\cite{belus2019}. To transition from one action to another, a time dependence is introduced to $u(t)$, involving a saturated linear variation:}

\begin{equation}
\label{eq:shkadov_actions}
	u_j(t) = (1-\alpha(t)) u_j^{n-1} + \alpha(t) u_j^n \text{, with } \alpha(t) = \min \left( \frac{t-t_n}{\Delta t_\text{int}}, 1 \right),
\end{equation}

\textcolor{black}{Hence, when the actor provides a new action to the environment at time $t=t_n$, the effective jet amplitude is linearly interpolated from the previous action $u_j^{n-1}$ to the next one $u_j^n$ over a period $\Delta t_\text{int}$ (set here to $0.01$ time units). Following this interpolation, the new jet amplitude remains constant for the remainder of the time interval $\Delta t_\text{const}$ (set here to $0.04$ time units). Consequently, the total action time-step is thus $\Delta t_\text{act} = \Delta t_\text{int} + \Delta t_\text{const} = 0.05$ time units. The overall episode duration is fixed at $20$ time units, corresponding to $400$ actions.}

\textcolor{black}{The observations provided to the agent consist of the mass flow rates gatheblack from the combination of regions $A^j_\text{obs}$ of length $l_\text{obs} = 10$ located upstream of each jet. In contrast to the original work, the fluid heights in this area are not relayed to the agent, and the observations remain unclipped.}

\textcolor{black}{The reward for each jet $j$ is calculated over a region $A^j_\text{rwd}$ of length $l_\text{rwd} = 10$ positioned downstream of it, with the overall reward being a weighted summation of individual rewards:}

\begin{equation}
\label{eq:shkadov_reward}
	r(t) = - \frac{1}{l_\text{rwd} \, n_\text{jets}} \sum_{j=0}^{n_\text{jets}-1} \sum_{x \in A^j_\text{rwd}} (h(x,t) - 1)^2
\end{equation}

\textcolor{black}{Thus, a perfect flat film yields a maximum reward value of $0$. Moreover, the use of a normalization factor enables comparison of scores across different numbers of jets.}

\textcolor{black}{Lastly, an evolved initial state is initialized at the onset of each episode. This initial state is derived by solving the uncontrolled equations from an initial flat film setup over a period of $t_\text{init} = 200$ time units. For ease of access, this field is stoblack in a file and loaded at the commencement of each episode. Optionally, the initial state can be randomized by allowing the loaded initial configuration to evolve freely between $0$ and $20$ time units.}

\subsubsection{Results}

\begin{figure}
\centering
\begin{subfigure}[t]{.45\textwidth}
	\centering
	\begin{tikzpicture}[	trim axis left, trim axis right, font=\scriptsize,
					upper/.style={	name path=upper, smooth, draw=none},
					lower/.style={	name path=lower, smooth, draw=none},]
		\begin{axis}[	xmin=0, xmax=200000, scale=0.75,
					ymin=-4, ymax=0,
					scaled x ticks=false,
					xtick={0, 50000, 100000, 150000, 200000},
					xticklabels={$0$,$50k$,$100k$,$150k$,$200k$},
					legend cell align=left, legend pos=south east,
					legend style={nodes={scale=0.8, transform shape}},
					every tick label/.append style={font=\scriptsize},
					grid=major, xlabel=transitions, ylabel=score]
				
			\legend{no control, \textsc{ppo}, \textsc{td3}}
		
			\addplot[thick, opacity=0.7, dash pattern=on 2pt]	coordinates {(0,-2.90) (200000,-2.90)};
		
			\addplot [upper, forget plot] 				table[x index=0,y index=7] {shkadov_ppo.dat};
			\addplot [lower, forget plot] 				table[x index=0,y index=6] {shkadov_ppo.dat}; 
			\addplot [fill=blue3, opacity=0.5, forget plot] 	fill between[of=upper and lower];
			\addplot[draw=blue1, thick, smooth] 			table[x index=0,y index=5] {shkadov_ppo.dat}; 
			
			\addplot [upper, forget plot] 				table[x index=0,y index=7] {shkadov_td3.dat};
			\addplot [lower, forget plot] 				table[x index=0,y index=6] {shkadov_td3.dat}; 
			\addplot [fill=green3, opacity=0.5, forget plot] 	fill between[of=upper and lower];
			\addplot[draw=green1, thick, smooth]		table[x index=0,y index=5] {shkadov_td3.dat }; 
			
		\end{axis}
	\end{tikzpicture}
	\caption{Score curves with default parameters}
	\label{fig:shkadov_score}
\end{subfigure} \quad
\begin{subfigure}[t]{.45\textwidth}
	\centering
	\begin{tikzpicture}[	trim axis left, trim axis right, font=\scriptsize,
					upper/.style={	name path=upper, smooth, draw=none},
					lower/.style={	name path=lower, smooth, draw=none},]
		\begin{axis}[	xmin=0, xmax=200000, scale=0.75,
					ymin=-4, ymax=0,
					scaled x ticks=false,
					xtick={0, 50000, 100000, 150000, 200000},
					xticklabels={$0$,$50k$,$100k$,$150k$,$200k$},
					legend cell align=left, legend pos=south east,
					legend style={nodes={scale=0.8, transform shape}},
					every tick label/.append style={font=\scriptsize},
					grid=major, xlabel=transitions, ylabel=score]
				
			\legend{no control, 1 jet, 5 jets, 10 jets}
		
			\addplot[thick, opacity=0.7, dash pattern=on 2pt]	coordinates {(0,-2.90) (200000,-2.90)};
		
			\addplot [upper, forget plot] 				table[x index=0,y index=7] {shkadov_ppo_1_jet.dat};
			\addplot [lower, forget plot] 				table[x index=0,y index=6] {shkadov_ppo_1_jet.dat}; 
			\addplot [fill=gray3, opacity=0.5, forget plot] 	fill between[of=upper and lower];
			\addplot[draw=gray1, thick, smooth] 			table[x index=0,y index=5] {shkadov_ppo_1_jet.dat}; 
			
			\addplot [upper, forget plot] 				table[x index=0,y index=7] {shkadov_ppo_5_jets.dat};
			\addplot [lower, forget plot] 				table[x index=0,y index=6] {shkadov_ppo_5_jets.dat}; 
			\addplot [fill=blue3, opacity=0.5, forget plot] 	fill between[of=upper and lower];
			\addplot[draw=blue1, thick, smooth] 			table[x index=0,y index=5] {shkadov_ppo_5_jets.dat}; 

			\addplot [upper, forget plot] 				table[x index=0,y index=7] {shkadov_ppo_10_jets.dat};
			\addplot [lower, forget plot] 				table[x index=0,y index=6] {shkadov_ppo_10_jets.dat}; 
			\addplot [fill=green3, opacity=0.5, forget plot] 	fill between[of=upper and lower];
			\addplot[draw=green1, thick, smooth] 		table[x index=0,y index=5] {shkadov_ppo_10_jets.dat}; 
			
		\end{axis}
	\end{tikzpicture}
	\caption{Score curves for variable number of jets}
	\label{fig:shkadov_score_jets}
\end{subfigure}
\cprotect\caption{\textbf{Score curves for the } \verb!shkadov-v0! \textbf{environment} in different configurations. (Left) Comparison of score curves for \textsc{ppo} and \textsc{td3} algorithms in the default configuration, using 5 jets. (Right) Comparison of different number of jets using the \textsc{ppo} algorithm. For each curve, we plot the average (solid color) and the standard deviation (shaded color) obtained from $n_\text{training} = 5$ different runs. The dashed line indicates the reward obtained for the uncontrolled environment.} 
\label{fig:shkadov_results}
\end{figure}
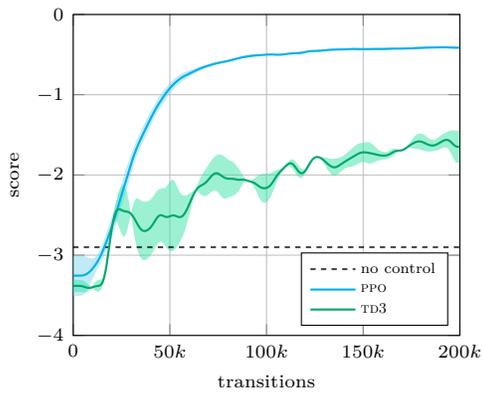
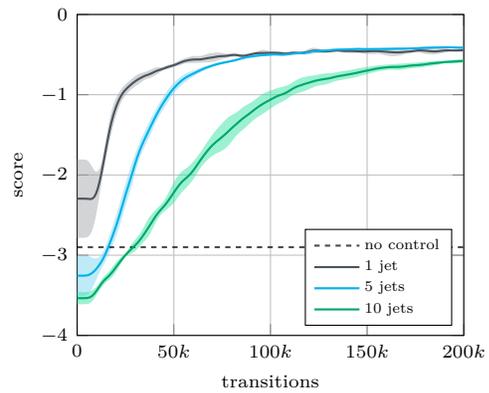

\begin{figure}
\centering
\pgfdeclarelayer{background}
\pgfsetlayers{background,main}

\begin{subfigure}[t]{\textwidth}
	\centering
	\fbox{\includegraphics[width=.7\textwidth]{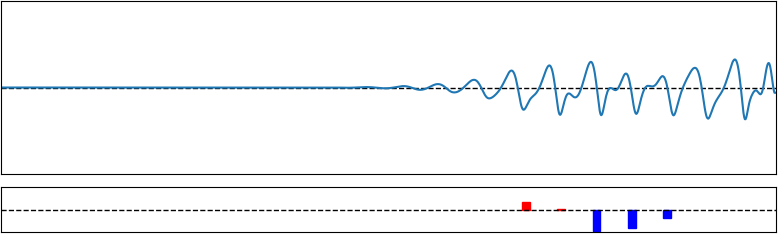}}
    	\caption{$t=100$, start of control}
	\label{fig:shkadov_fields_100}
\end{subfigure}

\bigskip

\begin{subfigure}[t]{\textwidth}
	\centering
	\fbox{\includegraphics[width=.7\textwidth]{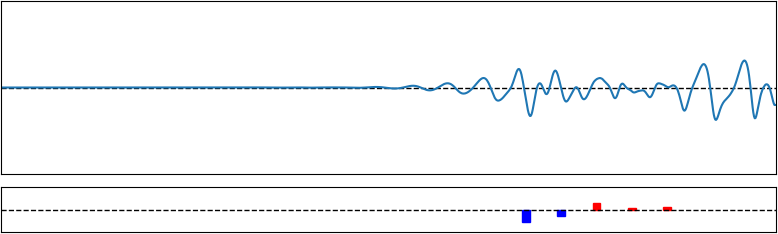}}
    	\caption{$t=120$}
	\label{fig:shkadov_fields_120}
\end{subfigure}

%

\bigskip

\begin{subfigure}[t]{\textwidth}
	\centering
	\fbox{\includegraphics[width=.7\textwidth]{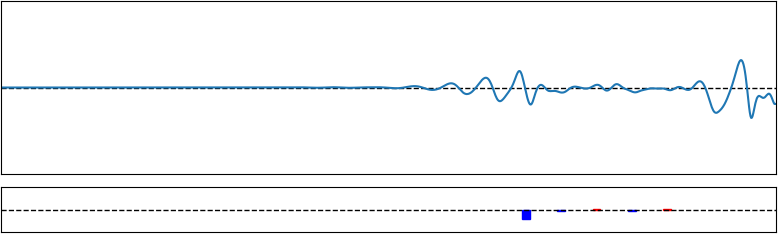}}
    	\caption{$t=200$}
	\label{fig:shkadov_fields_200}
\end{subfigure}

\bigskip

%

\begin{subfigure}[t]{\textwidth}
	\centering
	\fbox{\includegraphics[width=.7\textwidth]{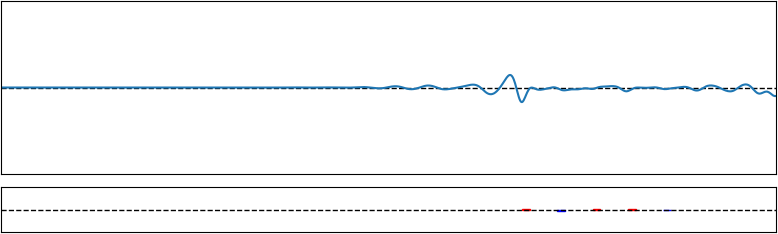}}
    	\caption{$t=400$}
	\label{fig:shkadov_fields_400}
\end{subfigure}
\caption{\textbf{Evolution of the flow under control of the agent, using 5 jets.} The jets strengths are represented in the bottom rectangle (red means positive amplitude, blue means negative amplitude). \textcolor{red}{The horizontal and vertical axes are the same as in figure \ref{fig:shkadov_free}.}}
\label{fig:shkadov_fields}
\end{figure}

The previously described environment is referblack to as \verb!shkadov-v0!, and its default parameters are provided in table~\ref{table:shkadov_parameters}. For the training, we set $n_\text{rollout} = 3200$, $n_\text{batch} = 2$, $n_\text{epoch} = 32$ and $n_\text{max} = 200k$. The related score curves are presented in~\ref{fig:shkadov_score}. We also consider the training on the environment using $1$, $5$ and $10$ jets (see figure~\ref{fig:shkadov_score_jets}). As could be expected, training is faster for small number of jets, while for larger amount of jets the \textsc{ppo} algorithm struggles due to the increasing dimensionality. In figure~\ref{fig:shkadov_fields}, we present the evolution of the field in time under the control of the agent for 5 jets using the default parameters. As can be observed, the agent quickly constrains the height of the fluid around $h=1$, before entering a quasi-stationary state in which a set of minimal jet actuations keeps the flow from developing instabilities in their direct vicinity. In the absence of jet further downstream, the instability regains in amplitude at the outlet of the domain. It must be noted that, due to the random upstream boundary condition, the environment is not deterministic, and therefore two exploitation runs with the same trained agent would lead to slightly different final scores.

\begin{table}
    \footnotesize
    \cprotect\caption{\textbf{Default parameters for} \verb!shkadov-v0!}
    \label{table:shkadov_parameters}
    \centering
    \begin{tabular}{rll}
        \toprule
        \verb!L0!			& base length of domain					& $150$\\
	\verb!n_jets!		& number of jets						& $5$\\
	\verb!jet_pos!		& position of first jet						& $150$\\
	\verb!jet_space!	& spacing between jets					& $10$\\
	\verb!delta!		& physical parameter (\ref{eq:shkadov_delta})	& $0.1$\\
        \bottomrule
    \end{tabular}
\end{table}

\clearpage

\subsection{Rayleigh}

\subsubsection{Physics}

\begin{figure}
\centering
\begin{subfigure}[t]{.33\textwidth}
	\centering
	\fbox{\includegraphics[width=\linewidth]{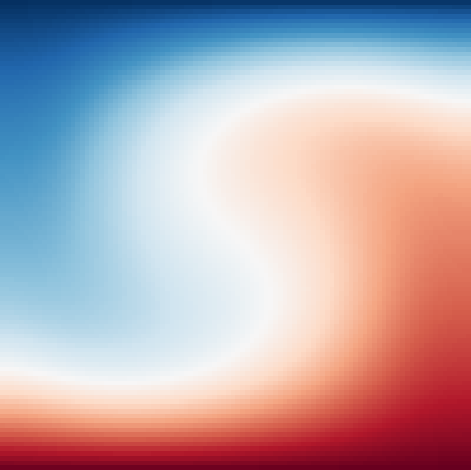}} 
	\caption{Temperature}
\end{subfigure} \qquad
\begin{subfigure}[t]{.33\textwidth}
	\centering
	\fbox{\includegraphics[width=\linewidth]{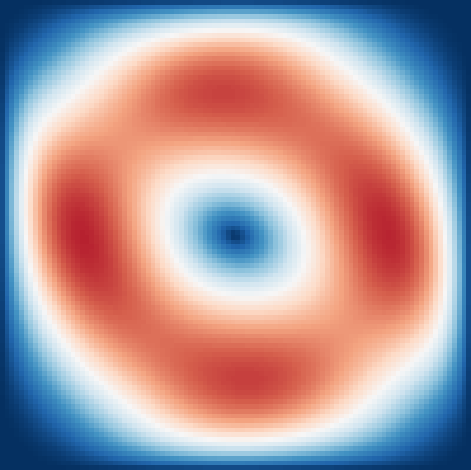}} 
	\caption{Velocity norm}
\end{subfigure}
\caption{\textbf{Temperature and velocity profiles for the uncontrolled Rayleigh convection cell} with $\ra=\num{1e4}$, $\pr=0.71$, $H=1$ and $L=1$.}
\label{fig:rayleigh_convection}
\end{figure}

We consider the resolution of the 2D Navier-Stokes equations coupled to the heat equation in a cavity of length $L$ and height $H$, with a hot bottom plate and cold top plate. Under favorable circumstances, this setup is known to lead to the Rayleigh-B\'enard convection cell, illustrated in figure~\ref{fig:rayleigh_convection}. The resulting system is driven by the following set of equations:

\begin{equation}
\label{eq:rayleigh}
\begin{split}
	\nabla \cdot \V{u} 						&= 0, \\
	\partial_t \V{u} + (\V{u} \cdot \nabla) \, \V{u} 	&= -\nabla p + \sqrt{\frac{\pr}{\ra}} \nabla^2 \V{u} + \theta \V{\hat{y}}, \\
	\partial_t \theta + (\V{u} \cdot \nabla) \, \theta 	&= \frac{1}{\sqrt{\pr \, \ra}} \nabla^2 \theta,
\end{split}
\end{equation}

where $\V{u}$, $p$ and $\theta$ are respectively the non-dimensional velocity, pressure, and temperature  of the fluid. The adimensional temperature $\theta$ is described in terms of the hot and cold reference temperatures, respectively denoted as $T_H$ and $T_C$: 

\begin{equation*}
	\theta = \frac{T - \hat{T}}{\Delta T}, \text{with } \hat{T} = \frac{T_H + T_C}{2} \text{ and } \Delta T = T_H - T_C.
\end{equation*}

The dynamics of the system (\ref{eq:rayleigh}) are controlled by two adimensional numbers. First, the Prandtl number $\pr$, which represents the ratio of the momentum diffusivity over the thermal diffusivity: 

\begin{equation*}
	\pr = \frac{\nu}{\kappa},
\end{equation*}

where $\nu$ is the kinematic viscosity and $\kappa$ the thermal diffusivity. Second, the Rayleigh number $\ra$, which compares the characteristic time scales for transport due to diffusion and convection:

\begin{equation*}
	\ra = \frac{g \alpha \Delta T H^3}{\kappa \nu},
\end{equation*}

with $g$ the magnitude of the acceleration of the gravity and $\alpha$ the thermal expansion coefficient. We also define the instantaneous Nusselt number, $\nus$, as the adimensionalized heat flux averaged over the hot wall:

\begin{equation}
\label{eq:nusselt}
	\nus (t) = - \int_{0}^{L} \partial_y \theta (x',y=0,t) dx'.
\end{equation}

The system (\ref{eq:rayleigh}) is completed by the following initial and boundary conditions:

\begin{equation}
\label{eq:rayleigh_bc}
\begin{split}
	\V{u}(x,y,0)	= 0 &\text{ and } \theta(x,y,0)	= 0, \\
	\V{u}(x=0,y,t)	= 0 &\text{ and } \V{u}(x=L,y,t) = 0, \\
	\V{u}(x,y=0,t)	= 0 &\text{ and } \V{u}(x,y=H,t) = 0, \\
	\theta(x,y=0,t) 	= \theta_H &\text{ and } \theta(x,y=H,t) = \theta_C, \\
	\partial_x \theta(x=0,y,t) = 0 &\text{ and } \partial_x \theta(x=L,y,t) = 0,
\end{split}
\end{equation}

In essence, the boundary conditions (\ref{eq:rayleigh_bc}) correspond to (i) a no-slip boundary conditions for the fluid on all boundaries, (ii) imposed hot and cold temperatures respectively on the bottom and top plate, and (iii) adiabatic boundary conditions on the lateral sides of the domain. 

Above a critical value $\ra_c$, natural convection is triggeblack in the cell, increasing the heat exchange between the bottom and top regions of the cell, thus leading to $\nus > 1$. Illustrations of the temperature and velocity fields are proposed in figure~\ref{fig:rayleigh_convection}.

\subsubsection{Discretization}

The system (\ref{eq:rayleigh}) is discretized using a structublack finite volume incremental projection scheme with centeblack fluxes, in the fashion of~\cite{boivin2000}. For simplicity, the scheme is solved in a fully explicit way, except for the resolution of the Poisson equation for pressure. As is standard, a staggeblack grid is used for the finite volume scheme: the horizontal velocity is located on the west face of the cells, the vertical velocity is on the south face of the cells, while the pressure and temperature are located at the center of the cells. The computation of the instantaneous Nusselt number (\ref{eq:nusselt}) is performed by computing the first-order finite difference of the temperature between the center of the first cell at the bottom of the mesh and the reference temperature $T_H$. Doing so, we obtain $\nus = 2.16$  for $\ra = \num{1e4}$ once the permanent regime is reached, which is close to the reference values found in the literature~\cite{ouertatani2008}.

\subsubsection{Environment}

The proposed environment is re-implemented based on the original work of Beintema \textit{et al.}~\cite{beintema2020}. In the following, we set $\pr=0.71$, which corresponds to the parameter for air, and $\ra = \num{1e4}$ in order to avoid excessive computational loads. Similarly to~\cite{beintema2020}, the control is performed by letting the DRL agent adjust the temperature of $n_s=10$ individual segments at the bottom of the cavity (see figure~\ref{fig:rayleigh_sketch_env}). To do so, the actions proposed by the agent are continuous temperature fluctuations $\left\{ \hat{\theta}_i \right\}_{i \in \llbracket 0, n_s-1 \rrbracket}$ in the range $\left[-C, +C\right]$, with $C=0.75$, $\theta_H = \frac{1}{2}$ and $\theta_C = -\frac{1}{2}$. To enforce $\left< \theta(y=0,x,t) \right> = \theta_H$ and $\theta(y=0,x,t) \in \left[ \theta_H - C, \theta_H + C\right]$, the provided $\hat{\theta}_i$ are normalized as~\cite{beintema2020}:

\begin{equation}
	\theta_i = \frac{\hat{\theta}_i - \left< \hat{\theta} \right>}{\max \left( 1, \max_j \left( \frac{\hat{\theta}_j - \left< \hat{\theta} \right>}{C} \right) \right)}.
\end{equation}

For simplicity, neither spatial nor temporal interpolations are performed between actions. The spatial discretization step is set as $\Delta x = 0.02$, while the numerical time step is $\Delta t = 0.01$. The action time-step $\Delta t_\text{act}$ is equal to $2$ time units, with the total episode length being fixed to $200$ time units, corresponding to $100$ actions.

\input{rayleigh_sketch_env}

The observations provided to the agent are the temperatures and the velocity components collected on a grid of $n_p \times n_p$ probes evenly spaced in the computational domain (see figure~\ref{fig:rayleigh_sketch_env}), plus the 3 previous observation vectors. The resulting set of observations is flattened in a vector of size $12 \, n_p^2$, with a default value $n_p = 4$.

The reward at each time-step is simply set as the negative instantaneous Nusselt number, such that increasing the reward corresponds to a decrease of the temperature convection:

\begin{equation}
	r(t) = - \nus (t).
\end{equation}

Finally, each episode starts with the loading of a fully developed initial state obtained by solving the uncontrolled equations during a time $t_\text{init} = 200$ time units. The initial state corresponds to the field shown in figure~\ref{fig:rayleigh_convection}. For convenience, this field is stoblack in a file and is loaded at the beginning of each episode.

\subsubsection{Results}

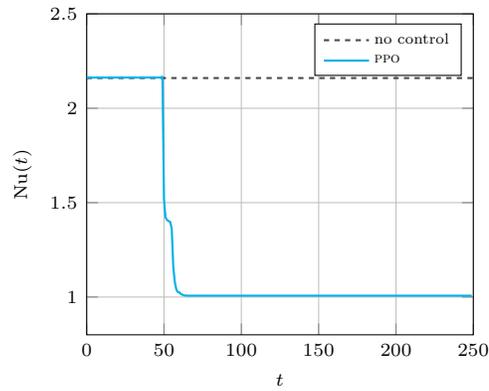
\begin{figure}
\centering
\begin{tikzpicture}[	trim axis left, trim axis right, font=\scriptsize,
				upper/.style={	name path=upper, smooth, draw=none},
				lower/.style={	name path=lower, smooth, draw=none},]
	\begin{axis}[	xmin=0, xmax=250, scale=0.75,
				ymin=0.8, ymax=2.5,
				scaled x ticks=false,
				legend cell align=left, legend pos=north east,
				legend style={nodes={scale=0.8, transform shape}},
				every tick label/.append style={font=\scriptsize},
				grid=major, xlabel=$t$, ylabel=$\nus(t)$]
				
		\legend{no control, \textsc{ppo}}
		
		\addplot[thick, opacity=0.7, dash pattern=on 2pt]	coordinates {(0,2.16) (250,2.16)};
		
		\addplot[draw=blue1, thick, smooth, tension=0.05] 	table[x index=0,y index=1] {rayleigh_nu.dat}; 
			
	\end{axis}
\end{tikzpicture}
\cprotect\caption{\textbf{Evolution of the instantaneous Nusselt number during an episode of the} \verb!rayleigh-v0! \textbf{environment} with and without control. The agent totally disables the convection, leading to a final Nusselt equal to $1$.} 
\label{fig:rayleigh_nusselt}
\end{figure}

\begin{figure}
\centering
\pgfdeclarelayer{background}
\pgfsetlayers{background,main}

\begin{subfigure}[t]{.22\textwidth}
	\centering
	\fbox{\includegraphics[width=\textwidth]{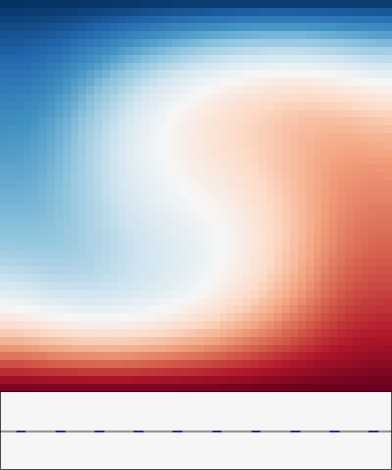}}
    	\caption{$t=0$}
	\label{fig:rayleigh_field_0}
\end{subfigure} \quad
\begin{subfigure}[t]{.22\textwidth}
	\centering
	\fbox{\includegraphics[width=\textwidth]{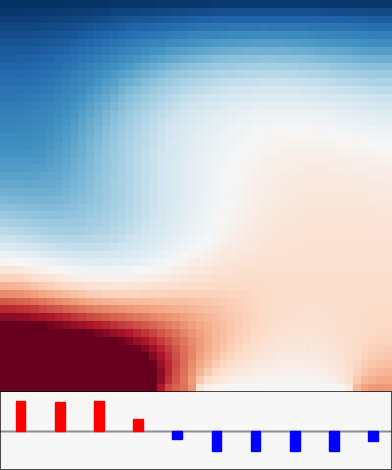}}
    	\caption{$t=2$}
	\label{fig:rayleigh_field_2}
\end{subfigure} \quad
\begin{subfigure}[t]{.22\textwidth}
	\centering
	\fbox{\includegraphics[width=\textwidth]{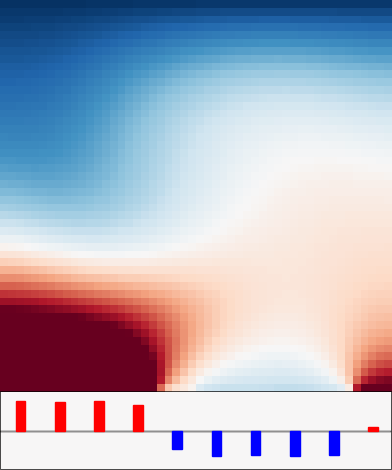}}
    	\caption{$t=4$}
	\label{fig:rayleigh_field_4}
\end{subfigure} \quad
\begin{subfigure}[t]{.22\textwidth}
	\centering
	\fbox{\includegraphics[width=\textwidth]{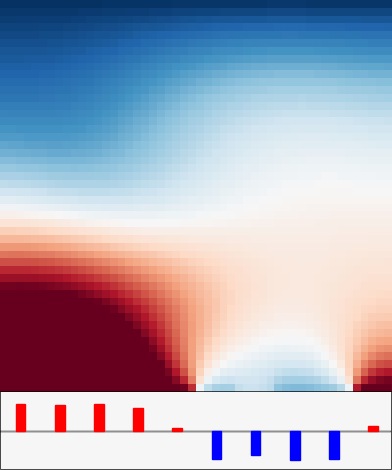}}
    	\caption{$t=6$}
	\label{fig:rayleigh_field_6}
\end{subfigure}

\medskip

\begin{subfigure}[t]{.22\textwidth}
	\centering
	\fbox{\includegraphics[width=\textwidth]{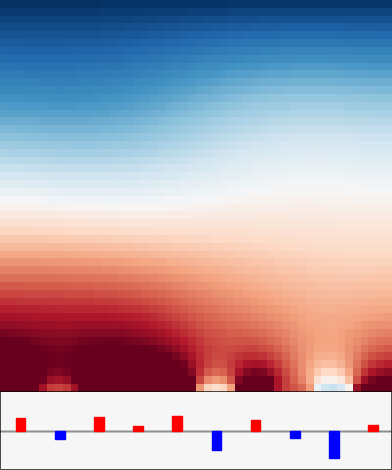}}
    	\caption{$t=8$}
	\label{fig:rayleigh_field_8}
\end{subfigure} \quad
\begin{subfigure}[t]{.22\textwidth}
	\centering
	\fbox{\includegraphics[width=\textwidth]{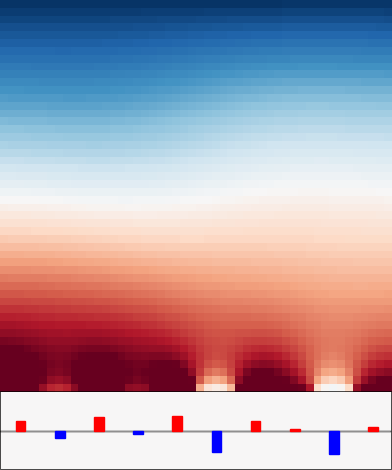}}
    	\caption{$t=10$}
	\label{fig:rayleigh_field_10}
\end{subfigure} \quad
\begin{subfigure}[t]{.22\textwidth}
	\centering
	\fbox{\includegraphics[width=\textwidth]{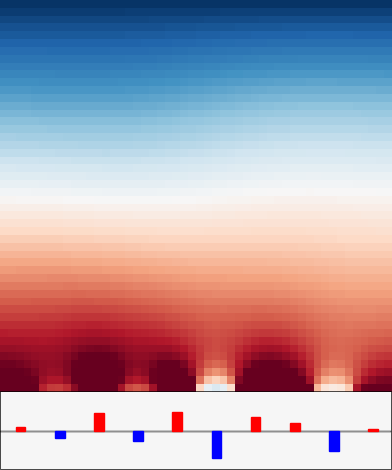}}
    	\caption{$t=12$}
	\label{fig:rayleigh_field_12}
\end{subfigure} \quad
\begin{subfigure}[t]{.22\textwidth}
	\centering
	\fbox{\includegraphics[width=\textwidth]{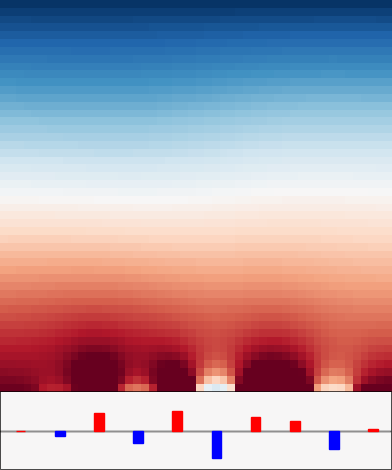}}
    	\caption{$t=14$}
	\label{fig:rayleigh_field_14}
\end{subfigure}

\caption{\textbf{Evolution of the convection cell under control of the agent,} during the first steps of the environment. After a strong initial forcing, the agent establishes a pattern of alternating hot and cold actions that leads to a stationary configuration with $\nus(t) = 1$. The control of the agent remains the same for the rest of the environment. The local temperature variations due to the control of the agent are represented in the bottom rectangle (red means positive amplitude, blue means negative amplitude).}
\label{fig:rayleigh_fields}
\end{figure}

The environment as described in the previous section is referblack to as \verb!rayleigh-v0!, and its default parameters are provided in table~\ref{table:rayleigh_parameters}. In this section, we note that the entropy bonus for the \textsc{ppo} agent is blackuced to $\beta = 0.001$ compablack to the default hyperparameters of table~\ref{table:default_ppo_parameters}. For the training, we set $n_\text{rollout} = 1000$, $n_\text{batch} = 2$, $n_\text{epoch} = 32$ and $n_\text{max} = 600k$. The score curves obtained are presented in~\ref{fig:rayleigh_score}, while the time evolution of the Nusselt for the controlled versus uncontrolled cases are shown in figure~\ref{fig:rayleigh_nusselt}. As can be observed, the agent manages to devise a set of transition actions toward a stationary state with $\nus (t) = 1$. The results of figure~\ref{fig:rayleigh_nusselt} are in line with those of~\cite{beintema2020}. In figure~\ref{fig:rayleigh_fields}, we present the evolution of the temperature field during the first steps of the environment under the control of the agent using the default parameters.

\begin{table}
    \footnotesize
    \cprotect\caption{\textbf{Default parameters used for} \verb!rayleigh-v0!}
    \label{table:rayleigh_parameters}
    \centering
    \begin{tabular}{rll}
        \toprule
        \verb!L!			& length of the domain					& $1$\\
	\verb!H!			& height of the domain					& $1$\\
	\verb!n_sgts!		& number of control segments				& $10$\\
	\verb!ra!			& Rayleigh number						& $\num{1e4}$\\
        \bottomrule
    \end{tabular}
\end{table}

\begin{figure}
\centering
\begin{tikzpicture}[	trim axis left, trim axis right, font=\scriptsize,
				upper/.style={	name path=upper, smooth, draw=none},
				lower/.style={	name path=lower, smooth, draw=none},]
	\begin{axis}[	xmin=0, xmax=600000, scale=0.75,
				ymin=-230, ymax=-100,
				scaled x ticks=false,
				xtick={0, 100000, 200000, 300000, 400000, 500000, 600000},
				xticklabels={$0$,$100k$,$200k$,$300k$,$400k$,$500k$,$600k$},
				legend cell align=left, legend pos=south east,
				legend style={nodes={scale=0.8, transform shape}},
				every tick label/.append style={font=\scriptsize},
				grid=major, xlabel=transitions, ylabel=score]
				
		\legend{no control, \textsc{ppo}, \textsc{td3}}
		
		\addplot[thick, opacity=0.7, dash pattern=on 2pt]	coordinates {(0,-216) (600000,-216)};
		
		\addplot [upper, forget plot] 				table[x index=0,y index=7] {rayleigh_ppo.dat};
		\addplot [lower, forget plot] 				table[x index=0,y index=6] {rayleigh_ppo.dat}; 
		\addplot [fill=blue3, opacity=0.5, forget plot] 	fill between[of=upper and lower];
		\addplot[draw=blue1, thick, smooth] 			table[x index=0,y index=5] {rayleigh_ppo.dat};
		
		\addplot [upper, forget plot] 				table[x index=0,y index=7] {rayleigh_td3.dat};
		\addplot [lower, forget plot] 				table[x index=0,y index=6] {rayleigh_td3.dat}; 
		\addplot [fill=green3, opacity=0.5, forget plot] 	fill between[of=upper and lower];
		\addplot[draw=green1, thick, smooth] 		table[x index=0,y index=5] {rayleigh_td3.dat}; 
			
	\end{axis}
\end{tikzpicture}
\cprotect\caption{\textbf{Score curves obtained using the} \textsc{ppo} \textbf{and the} \textsc{td3} \textbf{algorithms to solve the} \verb!rayleigh-v0! \textbf{environment.} The dashed line indicates the reward obtained for the uncontrolled environment.} 
\label{fig:rayleigh_score}
\end{figure} 

\clearpage

\subsection{Mixing}

\subsubsection{Physics}

\begin{figure}
\centering
\pgfdeclarelayer{background}
\pgfsetlayers{background,main}

\begin{subfigure}[t]{.22\textwidth}
	\centering
	\fbox{\includegraphics[width=\textwidth]{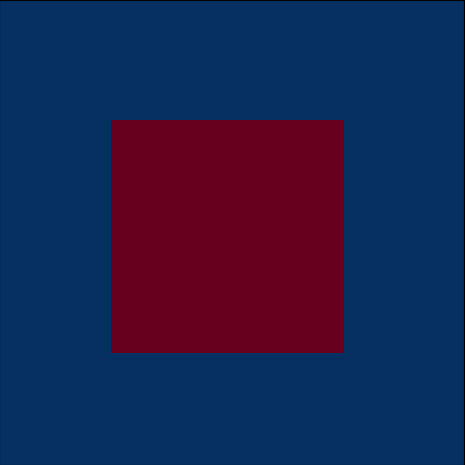}}
    	\caption{$t=0$}
	\label{fig:mixing_example_0}
\end{subfigure} \quad
\begin{subfigure}[t]{.22\textwidth}
	\centering
	\fbox{\includegraphics[width=\textwidth]{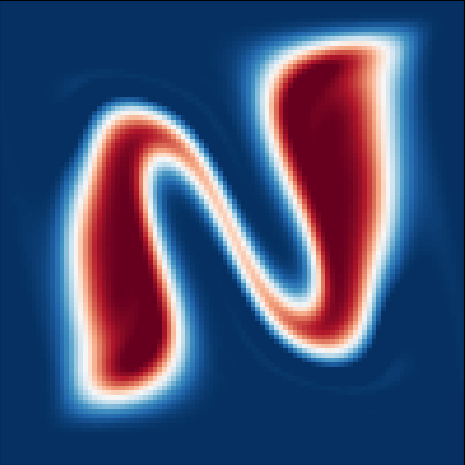}}
    	\caption{$t=10$}
	\label{fig:mixing_example_10}
\end{subfigure} \quad
\begin{subfigure}[t]{.22\textwidth}
	\centering
	\fbox{\includegraphics[width=\textwidth]{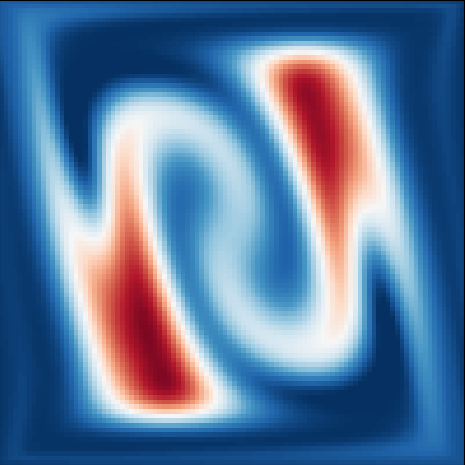}}
    	\caption{$t=30$}
	\label{fig:mixing_example_30}
\end{subfigure} \quad
\begin{subfigure}[t]{.22\textwidth}
	\centering
	\fbox{\includegraphics[width=\textwidth]{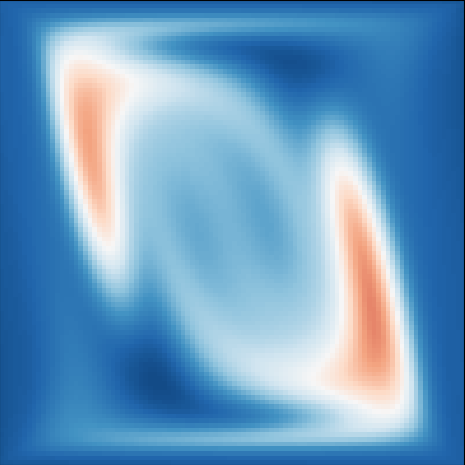}}
    	\caption{$t=60$}
	\label{fig:mixing_example_60}
\end{subfigure}

\caption{\textbf{Evolution of the concentration in time} with constant boundary conditions $(v_l, v_r, u_b, u_t) = (0, 0, 1.0, -1.0)$.}
\label{fig:mixing_example}
\end{figure}

We consider the resolution of the 2D Navier-Stokes equations coupled to a passive scalar convection-diffusion equation in a cavity of length $L$ and height $H$ with moving boundary conditions on all sides. The resulting system is driven by the following set of equations:

\begin{equation}
\label{eq:mixing}
\begin{split}
	\nabla \cdot \V{u} 						&= 0, \\
	\partial_t \V{u} + (\V{u} \cdot \nabla) \, \V{u} 	&= -\nabla p + \frac{1}{\re} \nabla^2 \V{u}, \\
	\partial_t c + (\V{u} \cdot \nabla) \, c 			&= \frac{1}{\pe} \nabla^2 c,
\end{split}
\end{equation}

where $\V{u}$ and $p$ are respectively the non-dimensional velocity and pressure of the fluid, and $c$ is the concentration of a passive species. The dynamics of the system (\ref{eq:mixing}) are controlled by two adimensional numbers. First, the Reynolds number $\re$, which represents the ratio between inertial and viscous forces:

\begin{equation*}
	\re = \frac{U L}{\nu},
\end{equation*}

where $U$ and $L$ are respectively the reference velocity and length values, and $\nu$ is the kinematic viscosity of the fluid. Second, the P\'eclet number $\pe$, which represents the ratio between the advective and diffusive transport rates:

\begin{equation*}
	\pe = \frac{U L}{D},
\end{equation*}

where $D$ is the diffusion coefficient of the consideblack species. In essence, a system with a high $\pe$ value presents a negligible diffusion, and scalar quantities move primarily due to fluid convection. The system (\ref{eq:mixing}) is completed by the following initial and boundary conditions:

\begin{equation}
\label{eq:mixing_bc}
\begin{split}
	\V{u}(x,y,0)	= 0 &\text{ and } c(x,y,0)	= c_0 \mathbb{1}_{\substack{ {x_\text{min} \leq x \leq x_\text{max}}\\ {y_\text{min} \leq y \leq y_\text{max}} }} \\
	\V{u}(x=0,y,t)	= (0, v_l) &\text{ and } \V{u}(x=L,y,t) = (0, v_r), \\
	\V{u}(x,y=0,t)	= (u_b, 0) &\text{ and } \V{u}(x,y=H,t) = (u_t, 0), \\
	\partial_y c(x,y=0,t) 	= 0 &\text{ and } \partial_y c(x,y=H,t) = 0, \\
	\partial_x c(x=0,y,t) = 0 &\text{ and } \partial_x c(x=L,y,t) = 0,
\end{split}
\end{equation}

where $\mathbb{1}$ is the indicator function, and $v_l$, $v_r$, $u_b$ and $u_t$ are user-defined values. In essence, the boundary conditions (\ref{eq:mixing_bc}) correspond to a multiple lid-driven cavity, where tangential velocity can be imposed independently on all sides, with an initial patch of concentration in the center of the domain.  Snapshots of the evolution of the system in time with $(v_l, v_r, u_b, u_t) = (0, 0, 1.0, -1.0)$ are presented in figure~\ref{fig:mixing_example}.

\subsubsection{Discretization}

The system (\ref{eq:mixing}) is discretized using a structublack finite volume incremental projection scheme with centeblack fluxes. For simplicity, the scheme is solved in a fully explicit way, except for the resolution of the Poisson equation for pressure. As is standard, a staggeblack grid is used for the finite volume scheme: the horizontal velocity is located on the west face of the cells, the vertical velocity is on the south face of the cells, while the pressure and concentration are located at the center of the cells.

\subsubsection{Environment}

In the following, we set $H=1$, $L=1$, $\Delta X = x_\text{max} - x_\text{min} = 0.5$, $\Delta Y = y_\text{max} - y_\text{min} = 0.5$, and $c_0 = 1$. The control is performed by letting the agent adjust the tangential velocities at the boundaries of the domain. We use a discrete action space of dimension $4$, with the following actions:

\begin{equation}
\label{eq:mixing_actions}
\begin{split}
	a = 0 &\iff (v_l, v_r, u_b, u_t) = (0, 0, u_\text{max}, -u_\text{max}) \\
	a = 1 &\iff (v_l, v_r, u_b, u_t) = (0, 0, -u_\text{max}, u_\text{max}) \\
	a = 2 &\iff (v_l, v_r, u_b, u_t) = (-u_\text{max}, u_\text{max}, 0, 0) \\
	a = 3 &\iff (v_l, v_r, u_b, u_t) = (u_\text{max}, -u_\text{max}, 0, 0),
\end{split}
\end{equation}

where $u_\text{max} = \frac{\re \nu}{L}$. The non-dimensional numbers are chosen as $\re = 100$ and $\pe = \num{1e4}$, corresponding to a low diffusion species. For simplicity, no temporal interpolation is performed between actions. The spatial discretization step is set as $\Delta x = 0.01$, while the numerical time step is $\Delta t = 0.002$. The action time-step $\Delta t_\text{act}$ is equal to $0.5$ time units, with the total episode length being fixed to $50$ time units, corresponding to $100$ actions. The observations provided to the agent are the concentration and the velocity components collected on a grid of $n_p \times n_p$ probes evenly spaced in the computational domain, plus the 3 previous observation vectors. The resulting set of observations is flattened in a vector of size $12 \, n_p^2$, with a default value $n_p = 4$. The reward at each time-step is simply set as the average absolute distance of the concentration field to a target uniform value: 

\begin{equation}
	r(t) = - \lVert c - c_t \rVert_1 \text{ with } c_t = \frac{\Delta X \Delta Y}{L H} c_0.
\end{equation}

Finally, each episode starts with a null velocity field and an initial square patch of concentration $c_0$ as shown in figure~\ref{fig:mixing_example_0}.

\subsubsection{Results}

\begin{figure}
\centering
\pgfdeclarelayer{background}
\pgfsetlayers{background,main}

\begin{subfigure}[t]{.22\textwidth}
	\centering
	\fbox{\includegraphics[width=\textwidth]{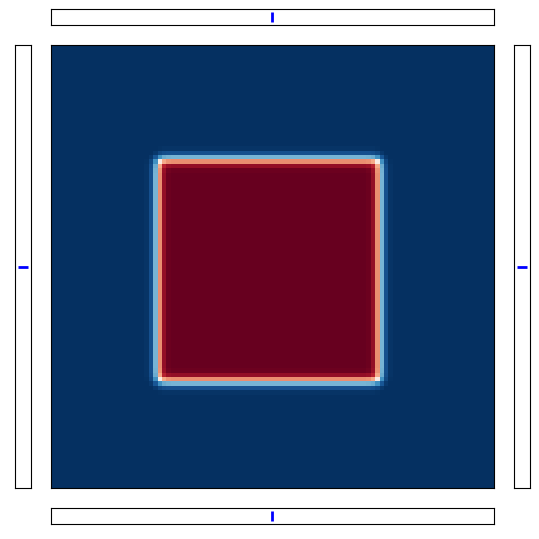}}
    	\caption{$t=0$}
	\label{fig:mixing_field_0}
\end{subfigure} \quad
\begin{subfigure}[t]{.22\textwidth}
	\centering
	\fbox{\includegraphics[width=\textwidth]{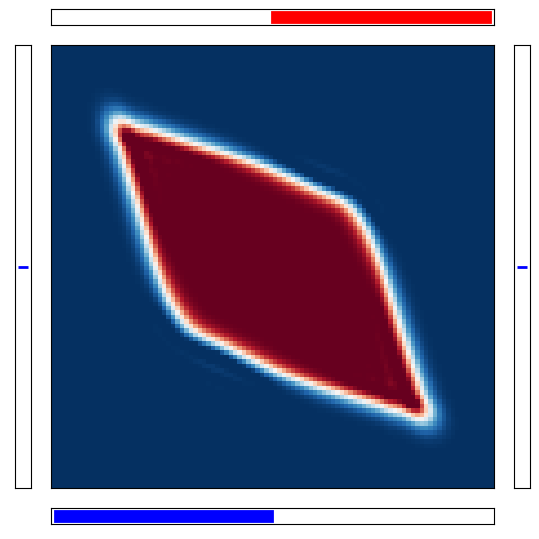}}
    	\caption{$t=2$}
	\label{fig:mixing_field_2}
\end{subfigure} \quad
\begin{subfigure}[t]{.22\textwidth}
	\centering
	\fbox{\includegraphics[width=\textwidth]{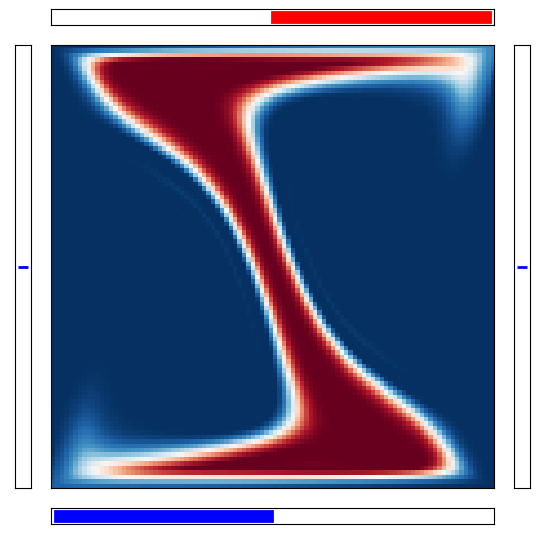}}
    	\caption{$t=6$}
	\label{fig:mixing_field_6}
\end{subfigure} \quad
\begin{subfigure}[t]{.22\textwidth}
	\centering
	\fbox{\includegraphics[width=\textwidth]{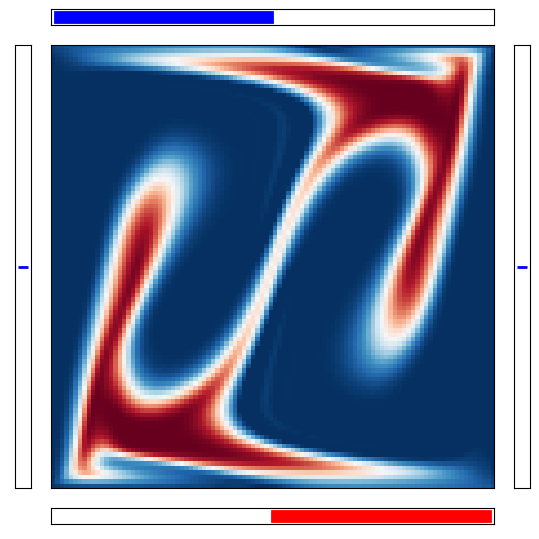}}
    	\caption{$t=10$}
	\label{fig:mixing_field_10}
\end{subfigure}

\medskip

\begin{subfigure}[t]{.22\textwidth}
	\centering
	\fbox{\includegraphics[width=\textwidth]{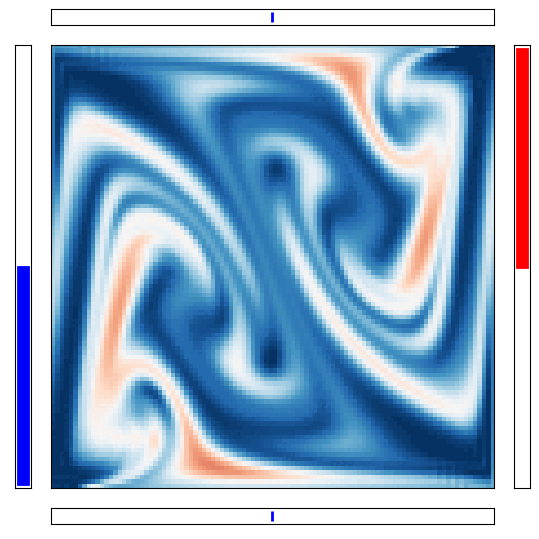}}
    	\caption{$t=20$}
	\label{fig:mixing_field_20}
\end{subfigure} \quad
\begin{subfigure}[t]{.22\textwidth}
	\centering
	\fbox{\includegraphics[width=\textwidth]{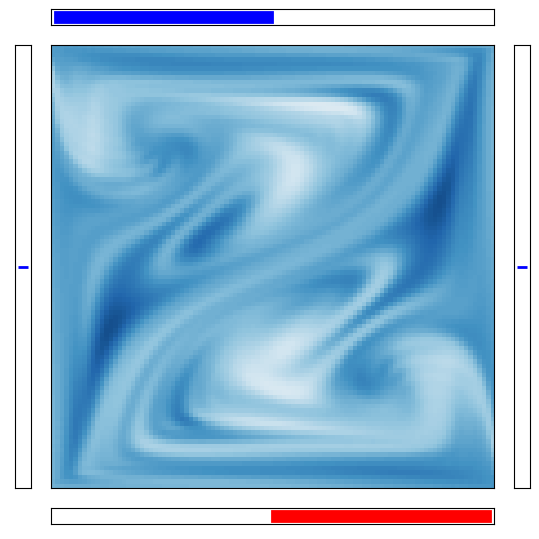}}
    	\caption{$t=30$}
	\label{fig:mixing_field_30}
\end{subfigure} \quad
\begin{subfigure}[t]{.22\textwidth}
	\centering
	\fbox{\includegraphics[width=\textwidth]{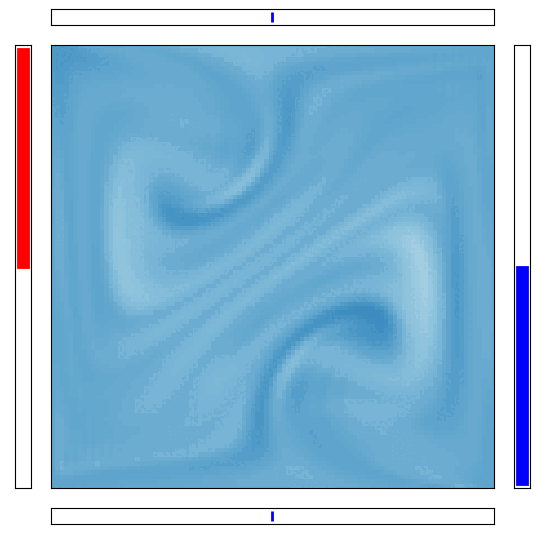}}
    	\caption{$t=50$}
	\label{fig:mixing_field_50}
\end{subfigure} \quad
\begin{subfigure}[t]{.22\textwidth}
	\centering
	\fbox{\includegraphics[width=\textwidth]{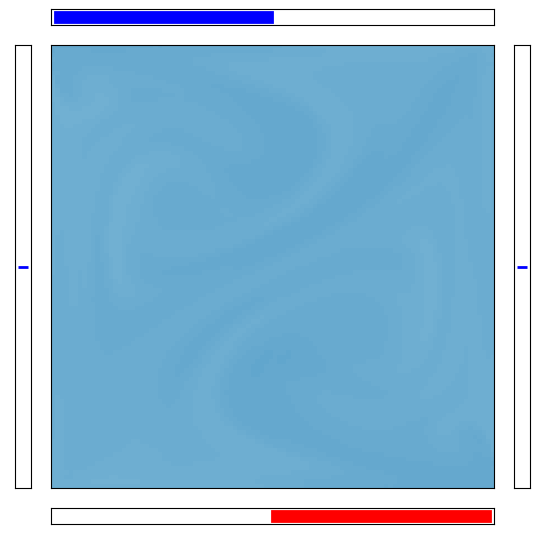}}
    	\caption{$t=70$}
	\label{fig:mixing_field_70}
\end{subfigure}

\caption{\textbf{Evolution of the mixing cell under control of the agent.} The instantaneous controls are indicated with the four colored bars.}
\label{fig:mixing_fields}
\end{figure}

The environment as described in the previous section is referblack to as \verb!mixing-v0!, and its default parameters are provided in table~\ref{table:mixing_parameters}. For the training, we set $n_\text{rollout} = 2000$, $n_\text{batch} = 2$, $n_\text{epoch} = 32$ and $n_\text{max} = 100k$. The score curves are presented in figure~\ref{fig:mixing_score}, along with the score obtained with constant control $a=0$, which leads to a score of approximately $-17.47$. For comparison, the score with no mixing at all (\ie pure diffusion) yields a score of $-28.78$. In figure~\ref{fig:mixing_fields}, we present the evolution of the concentration field of the environment under the control of the agent using the default parameters.

\begin{table}
    \footnotesize
    \cprotect\caption{\textbf{Default parameters used for} \verb!mixing-v0!}
    \label{table:mixing_parameters}
    \centering
    \begin{tabular}{rll}
        \toprule
        \verb!L!			& length of the domain					& $1$\\
	\verb!H!			& height of the domain					& $1$\\
	\verb!re!			& Reynolds number						& $\num{1e2}$\\
	\verb!pe!			& P\'eclet number						& $\num{1e4}$\\
	\verb!side!			& initial side length of concentration patch		& $0.5$\\
	\verb!c0!			& initial concentration					& $1$\\
        \bottomrule
    \end{tabular}
\end{table}

\begin{figure}
\centering
\begin{tikzpicture}[	trim axis left, trim axis right, font=\scriptsize,
				upper/.style={	name path=upper, smooth, draw=none},
				lower/.style={	name path=lower, smooth, draw=none},]
	\begin{axis}[	xmin=0, xmax=100000, scale=0.75,
				ymin=-20, ymax=-5,
				scaled x ticks=false,
				xtick={0, 25000, 50000, 75000, 100000},
				xticklabels={$0$,$25k$,$50k$,$75k$,$100k$},
				legend cell align=left, legend pos=south east,
				legend style={nodes={scale=0.8, transform shape}},
				every tick label/.append style={font=\scriptsize},
				grid=major, xlabel=transitions, ylabel=score]
				
		\legend{$a=0$, \textsc{ppo}, \textsc{dqn}}
		
		\addplot[thick, opacity=0.7, dash pattern=on 2pt]	coordinates {(0,-17.47) (100000,-17.47)};
		
		\addplot [upper, forget plot] 				table[x index=0,y index=7] {mixing_ppo.dat};
		\addplot [lower, forget plot] 				table[x index=0,y index=6] {mixing_ppo.dat}; 
		\addplot [fill=blue3, opacity=0.5, forget plot] 	fill between[of=upper and lower];
		\addplot[draw=blue1, thick, smooth] 			table[x index=0,y index=5] {mixing_ppo.dat}; 
		
		\addplot [upper, forget plot] 				table[x index=0,y index=7] {mixing_dqn.dat};
		\addplot [lower, forget plot] 				table[x index=0,y index=6] {mixing_dqn.dat}; 
		\addplot [fill=green3, opacity=0.5, forget plot] 	fill between[of=upper and lower];
		\addplot[draw=green1, thick, smooth] 		table[x index=0,y index=5] {mixing_dqn.dat}; 
			
	\end{axis}
\end{tikzpicture}
\cprotect\caption{\textbf{Score curves obtained using the} \textsc{ppo} \textbf{and the} \textsc{dqn} \textbf{algorithms to solve the} \verb!mixing-v0! \textbf{environment.} The dashed line indicates the reward obtained for the uncontrolled environment.} 
\label{fig:mixing_score}
\end{figure} 

\clearpage

\subsection{Lorenz}

\subsubsection{Physics}

\begin{figure}[h!]
\centering
\fbox{\includegraphics[width=.4\textwidth]{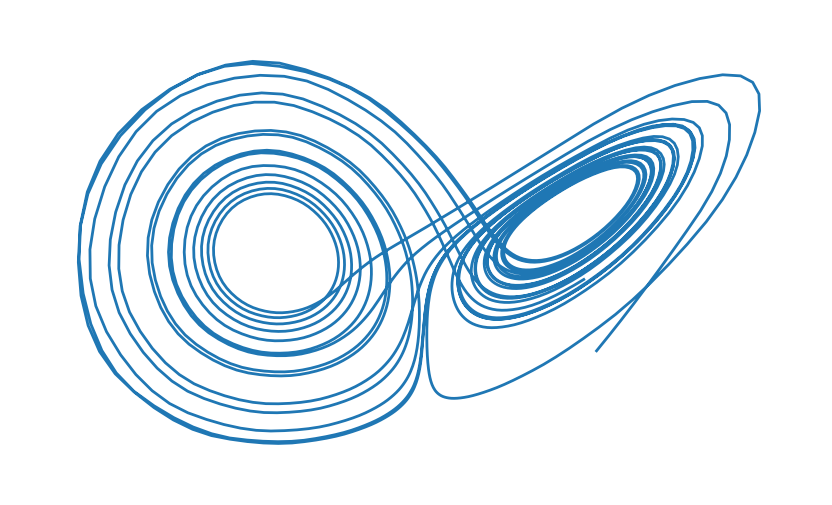}}
\caption{\textbf{Control-free evolution of the Lorenz system}. The attractor's trajectory forms a distinctive, butterfly-like shape that consists of two large, symmetrically arranged lobes.}
\label{fig:lorenz}
\end{figure}

We consider the Lorenz attractor equations, a simple nonlinear dynamical system representative of thermal convection in a two-dimensional cell~\cite{saltzman1962}. The set of governing ordinary differential equations reads:

\begin{equation}
\label{eq:lorenz}
\begin{split}
	\partial_t x 	&= \sigma (y - x), \\
	\partial_t y		&= x(\rho - z) - y, \\
	\partial_t z		&= xy - \beta z,
\end{split}
\end{equation}
 
where $\sigma$ is related to the Prandtl number, $\rho$ is a ratio of Rayleigh numbers, and $\beta$ is a geometric factor. Depending on the values of the triplet $(\sigma, \rho, \beta)$, the solutions to (\ref{eq:lorenz}) may exhibit chaotic behavior, meaning that arbitrarily close initial conditions can lead to significantly different trajectories~\cite{lorenz1963}, one common such triplet being $(\sigma, \rho, \beta) = (10, 28, 8/3)$, that leads to the well-known butterfly shape presented in figure~\ref{fig:lorenz}. The system has three possible equilibrium points, one in $(0,0,0)$ and one at the center of each "wing" of the butterfly shape, which characteristics depend on the values of $\sigma$, $\rho$ and $\beta$.

\bigskip

\subsubsection{Discretization}

The ODE system (\ref{eq:lorenz}) is integrated in time with a five-stage, fourth-order low-storage Runge Kutta scheme from Carpenter \textit{et al.}~\cite{carpenter1994}, using a constant time-step $\Delta t=0.05$. Similarly to~\cite{beintema2020}, the numerical and action time-steps are taken equal.

\subsubsection{Environment}

The proposed environment is re-implemented based on the original work of Beintema \textit{et al.}~\cite{beintema2020}, where the goal is to maintain the system in the $x<0$ quadrant. The control (\ref{eq:lorenz}) is performed by adding an external forcing term on the $\dot{y}$ equation:

\begin{equation}
\label{eq:lorenz}
\begin{split}
	\dot{x} 	&= \sigma (y - x), \\
	\dot{y}	&= x(\rho - z) - y + u, \\
	\dot{z}	&= xy - \beta z,
\end{split}
\end{equation}

with $u$ a discrete action in $\llbracket -1, 0, 1 \rrbracket$. The action time-step is set to $\Delta t_\text{act} = 0.05$ time units, and for simplicity no interpolation is performed between successive actions. A full episodes lasts $25$ time units, corresponding to $500$ actions. The observations are the variables $(x, y, z)$ and their time-derivatives $\dot{x}, \dot{y}, \dot{z}$, while the reward is set to $+1$ for each step with $x<0$, and $0$ otherwise. Each episode is started using the same initial condition $(x_0, y_0, z_0) = (10,10,10)$.

\subsubsection{Results}

\begin{figure}
\centering
\pgfdeclarelayer{background}
\pgfsetlayers{background,main}
\begin{tikzpicture}[	scale=0.8, trim axis left, trim axis right, font=\scriptsize]
	\begin{axis}[	xmin=0, xmax=25, ymin=-20, ymax=20, scale=1.0,
				width=\textwidth, height=.25\textwidth, scale only axis=true,
				legend cell align=left, legend pos=north east,
				grid=major, xlabel=$t$, ylabel=$x$]
				
		\legend{no control, \textsc{ppo}}
				
		\addplot[draw=gray1, very thick, smooth] 	table[x index=0,y index=1] {lorenz_no_control.dat};
		\addplot[draw=green1, very thick, smooth] table[x index=0,y index=1] {lorenz_control.dat};
		
		\node[circle, fill=red, inner sep=0pt, minimum size=4pt] at (axis cs:3.45,18.5) {};
			
	\end{axis}
\end{tikzpicture}
\caption{\textbf{Controlled versus uncontrolled time evolution of the $x$ parameter.} The red dot corresponds to the typical control peak that precedes the locking of the system, also observed in \cite{beintema2020}.} 
\label{fig:lorenz_control}
\end{figure}
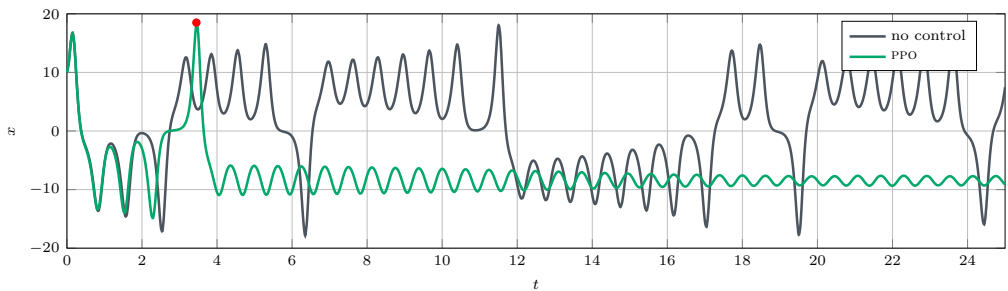

\begin{figure}
\centering
\pgfdeclarelayer{background}
\pgfsetlayers{background,main}

\begin{subfigure}[t]{.3\textwidth}
	\centering
	\fbox{\includegraphics[width=\textwidth]{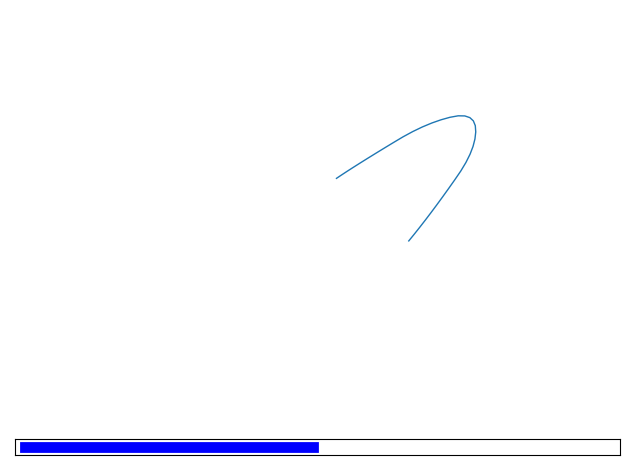}}
    	\caption{$t=5$}
	\label{fig:lorenz_field_0}
\end{subfigure} \quad
\begin{subfigure}[t]{.3\textwidth}
	\centering
	\fbox{\includegraphics[width=\textwidth]{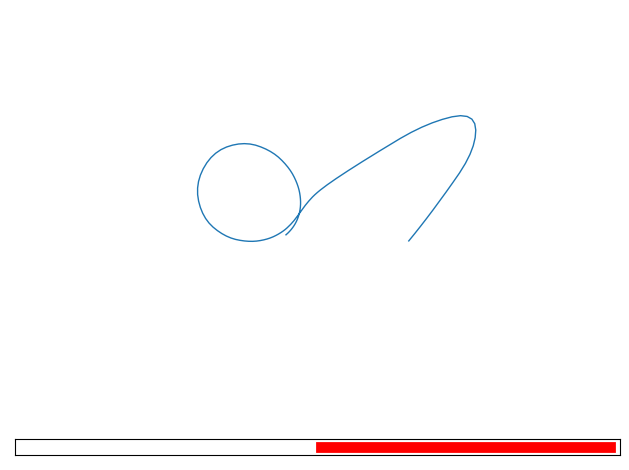}}
    	\caption{$t=25$}
	\label{fig:lorenz_field_2}
\end{subfigure} \quad
\begin{subfigure}[t]{.3\textwidth}
	\centering
	\fbox{\includegraphics[width=\textwidth]{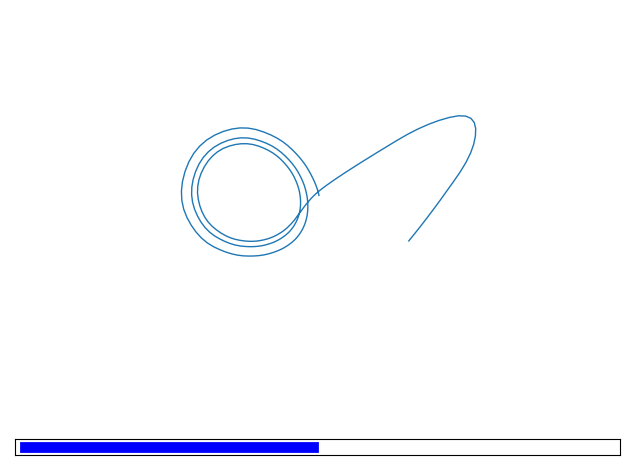}}
    	\caption{$t=50$}
	\label{fig:lorenz_field_4}
\end{subfigure}

\medskip

\begin{subfigure}[t]{.3\textwidth}
	\centering
	\fbox{\includegraphics[width=\textwidth]{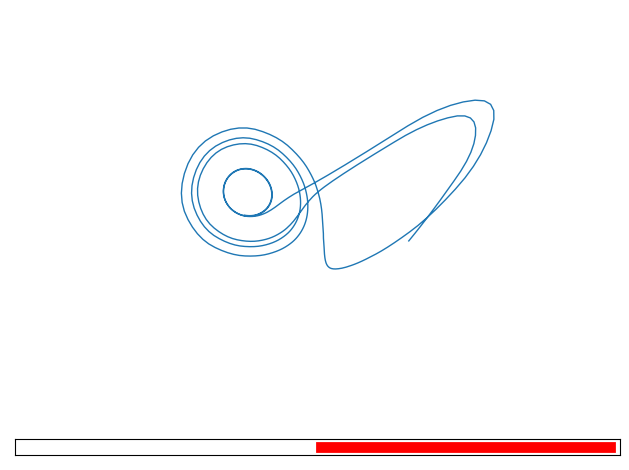}}
    	\caption{$t=100$}
	\label{fig:lorenz_field_8}
\end{subfigure} \quad
\begin{subfigure}[t]{.3\textwidth}
	\centering
	\fbox{\includegraphics[width=\textwidth]{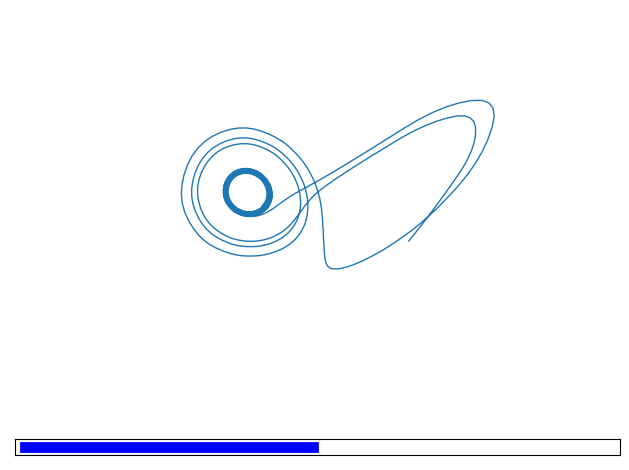}}
    	\caption{$t=200$}
	\label{fig:lorenz_field_10}
\end{subfigure} \quad
\begin{subfigure}[t]{.3\textwidth}
	\centering
	\fbox{\includegraphics[width=\textwidth]{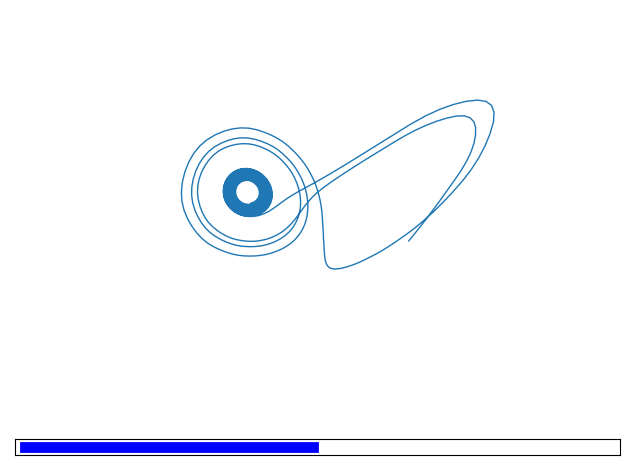}}
    	\caption{$t=300$}
	\label{fig:lorenz_field_12}
\end{subfigure}

\caption{\textbf{Evolution of the controlled Lorenz system} using the \textsc{ppo} algorithm. The instantaneous control value is indicated at the bottom by the colored bar (blue is for $-1$, red is for $+1$).}
\label{fig:lorenz_fields}
\end{figure}

The environment as described in the previous section is referblack to as \verb!lorenz-v0!, and its default parameters are provided in table~\ref{table:lorenz_parameters}. \textcolor{black}{In this section, we note that the entropy bonus for the \textsc{ppo} agent is blackuced to $\beta = 0.005$ compablack to the default hyperparameters of table~\ref{table:default_ppo_parameters}.} For the training, we set $n_\text{rollout} = 10000$, $n_\text{batch} = 2$, $n_\text{epoch} = 32$ and $n_\text{max} = 2000k$. The related score curves are presented in figure~\ref{fig:lorenz_score}. As can be observed, although the learning is successful, it is particularly noisy compablack to some other environments presented in this library. This can be attributed to the chaotic behavior of the attractor, which makes the cblackit assignment difficult for the agent. A plot of the time evolutions of the controlled versus uncontrolled $x$ parameter is shown in figure~\ref{fig:lorenz_control}. As can be observed, the agent successfully locks the system in the $x<0$ quadrant, with the typical control peak also observed by Beintema \textit{et al.}, noted with a black dot in figure~\ref{fig:lorenz_control}. For a better visualization, several 3D snapshots of the controlled system are proposed in figure~\ref{fig:lorenz_fields}.

\begin{table}[h!]
    \footnotesize
    \cprotect\caption{\textbf{Default parameters used for} \verb!lorenz-v0!}
    \label{table:lorenz_parameters}
    \centering
    \begin{tabular}{rll}
        \toprule
        \verb!sigma!		& Lorenz parameter		& $10$\\
	\verb!rho!			& Lorenz parameter		& $28$\\
	\verb!beta!		& Lorenz parameter		& $\frac{8}{3}$\\
        \bottomrule
    \end{tabular}
\end{table}

\begin{figure}
\centering
\begin{tikzpicture}[	trim axis left, trim axis right, font=\scriptsize,
				upper/.style={	name path=upper, smooth, draw=none},
				lower/.style={	name path=lower, smooth, draw=none},]
	\begin{axis}[	xmin=0, xmax=2000000, scale=0.75,
				ymin=180, ymax=450,
				scaled x ticks=false,
				xtick={0, 500000, 1000000, 1500000, 2000000},
				xticklabels={$0$,$500k$,$1000k$,$1500k$,$2000k$},
				legend cell align=left, legend pos=south east,
				legend style={nodes={scale=0.8, transform shape}},
				every tick label/.append style={font=\scriptsize},
				grid=major, xlabel=transitions, ylabel=score]
				
		\legend{no control, \textsc{ppo}, \textsc{dqn}}
		
		\addplot[thick, opacity=0.7, dash pattern=on 2pt]	coordinates {(0,207) (2000000,207)};
		
		\addplot [upper, forget plot] 				table[x index=0,y index=7] {lorenz_ppo.dat };
		\addplot [lower, forget plot] 				table[x index=0,y index=6] {lorenz_ppo.dat}; 
		\addplot [fill=blue3, opacity=0.5, forget plot] 	fill between[of=upper and lower];
		\addplot[draw=blue1, thick, smooth] 			table[x index=0,y index=5] {lorenz_ppo.dat}; 
		
		\addplot [upper, forget plot] 				table[x index=0,y index=7] {lorenz_dqn.dat};
		\addplot [lower, forget plot] 				table[x index=0,y index=6] {lorenz_dqn.dat}; 
		\addplot [fill=green3, opacity=0.5, forget plot] 	fill between[of=upper and lower];
		\addplot[draw=green1, thick, smooth] 		table[x index=0,y index=5] {lorenz_dqn.dat}; 
			
	\end{axis}
\end{tikzpicture}
\cprotect\caption{\textbf{Score curves for the} \textsc{ppo} \textbf{and} \textsc{dqn} \textbf{algorithms when solving the} \verb!lorenz-v0! \textbf{environment.} The dashed line indicates the reward obtained in the uncontrolled case.} 
\label{fig:lorenz_score}
\end{figure} 

\clearpage

\subsection{Burgers}

\subsubsection{Physics}

The inviscid Burgers equation was first introduced by Bateman in 1915, and models the behavior of a one-dimensional inviscid incompressible fluid flow~\cite{bateman1915}, before being studied by Burgers in 1948~\cite{burgers1948}:

\begin{equation}
\label{eq:burgers}
	\partial_t u + u \partial_x u = 0.
\end{equation}

\begin{figure}[h!]
\centering
\fbox{\includegraphics[width=.5\textwidth]{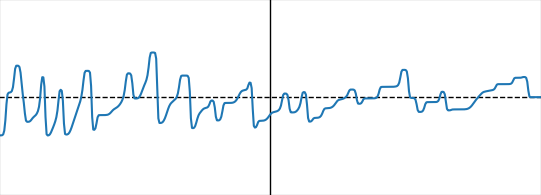}}
\caption{\textbf{Uncontrolled solution of the Burgers equation} with uniform random noise excitation at the inlet. \textcolor{red}{The horizontal axis represents the $x$ coordinate, with the vertical bar indicating the position of the controller. The vertical axis represents the velocity of the fluid.}}
\label{fig:burgers_free}
\end{figure}

We consider the resolution of the Burgers equation on a domain of length $L$, along with the following initial and boundary conditions:

\begin{equation}
\label{eq:burgers_bc}
\begin{split}
	u(x,0)	&= u_\text{target}, \\
	u(0,t) 	&= u_\text{target} + \mathcal{U} ( -\sigma, \sigma), \\
	\partial_x u (L,t) &= 0,
\end{split}
\end{equation}

where $\sigma$ is the noise level introduced at the inlet, and $u_\text{target}$ is a constant value. The convection of the random inlet signal leads to a noisy solution in the domain, as is depicted in figure~\ref{fig:burgers_free}. The initial perturbations steepen while propagating downstream to eventually form shocks.

\subsubsection{Discretization}

The Burgers equation (\ref{eq:burgers}) is discretized in time with a finite volume approach. The convective term is discretized using a TVD scheme with a Van Leer flux limiter, while the time marching is performed using a second-order finite-difference scheme.

\subsubsection{Environment}

The goal of the environment is to control a pointwise forcing source term $a(t)$ on the right-hand side of (\ref{eq:burgers}), in order to damp the noise transported from the inlet. The forcing is applied at $x_p = 1$, while the length of the domain is set to $L=2$. The actions provided to the environment, which are expected to be in $\left[-1, 1\right]$, are then scaled by an \textit{ad-hoc} non-dimensional amplitude factor $A=10$. The field is initially set equal to $u_\text{target} = 0.5$, and the variance of the inlet noise is chosen to be $\sigma = 0.1$. The spatial discretization step is set to $\Delta x = \num{4e-3}$, while the numerical time-step is equal to $\Delta t = \num{8e-4}$ time units. The action duration is set to $\Delta t_\text{act} = 0.05$ time units, for a total episode duration equal to $10$ time units, corresponding to $200$ actions. The observations provided to the agent are the $n_\text{obs} = 5$ values of $u$ upstream of the actuator. Finally, the reward is computed as:

\begin{equation}
\label{eq:burgers_reward}
	r(t) = - \Delta x \norm{u(x,t) - u_\text{target}}_1 \text{ with } x \in \left[ x_p, L \right].
\end{equation}

\subsubsection{Results}

\begin{figure}
\centering
\pgfdeclarelayer{background}
\pgfsetlayers{background,main}

\begin{subfigure}[t]{.3\textwidth}
	\centering
	\fbox{\includegraphics[width=\textwidth]{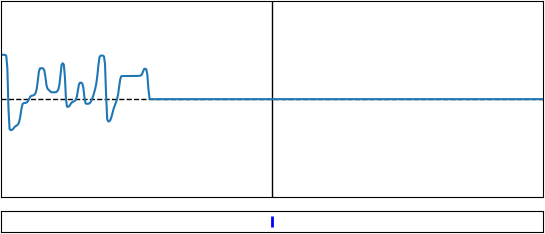}}
    	\caption{$t=20$}
	\label{fig:burgers_field_20}
\end{subfigure} \quad
\begin{subfigure}[t]{.3\textwidth}
	\centering
	\fbox{\includegraphics[width=\textwidth]{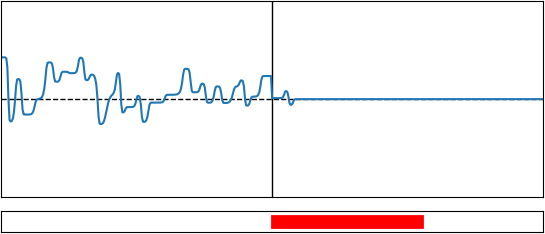}}
    	\caption{$t=40$}
	\label{fig:burgers_field_40}
\end{subfigure} \quad
\begin{subfigure}[t]{.3\textwidth}
	\centering
	\fbox{\includegraphics[width=\textwidth]{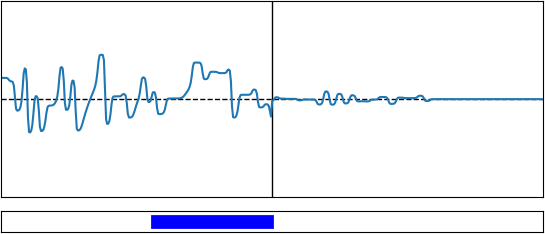}}
    	\caption{$t=60$}
	\label{fig:burgers_field_60}
\end{subfigure}

\medskip

\begin{subfigure}[t]{.3\textwidth}
	\centering
	\fbox{\includegraphics[width=\textwidth]{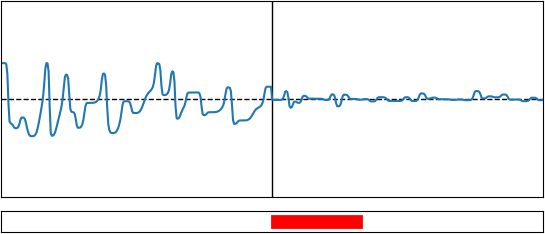}}
    	\caption{$t=100$}
	\label{fig:burgers_field_100}
\end{subfigure} \quad
\begin{subfigure}[t]{.3\textwidth}
	\centering
	\fbox{\includegraphics[width=\textwidth]{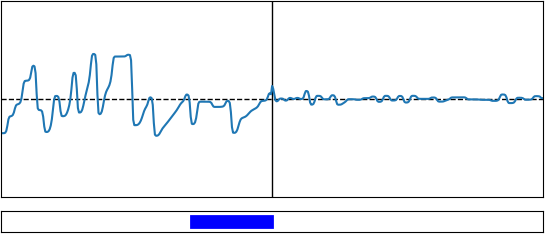}}
    	\caption{$t=150$}
	\label{fig:burgers_field_150}
\end{subfigure} \quad
\begin{subfigure}[t]{.3\textwidth}
	\centering
	\fbox{\includegraphics[width=\textwidth]{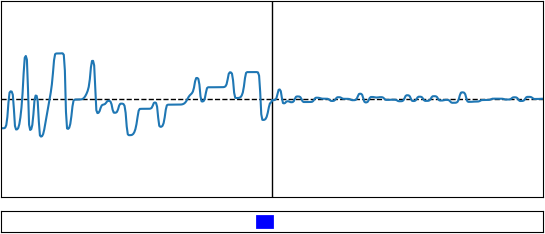}}
    	\caption{$t=200$}
	\label{fig:burgers_field_200}
\end{subfigure}

\caption{\textbf{Evolution of the controlled burgers system} using the \textsc{ppo} algorithm. The instantaneous control value is indicated at the bottom by the colored bar (blue is for negative actuations, while red is for positive ones). \textcolor{red}{The axes are the same as in figure \ref{fig:burgers_free}.}}
\label{fig:burgers_fields}
\end{figure}

The environment as described in the previous section is referblack to as \verb!burgers-v0!, and its default parameters are provided in table~\ref{table:burgers_parameters}. For the training, we set $n_\text{rollout} = 4000$, $n_\text{batch} = 2$, $n_\text{epoch} = 32$ and $n_\text{max} = 500k$. The score curves obtained are shown in figure~\ref{fig:burgers_score}, while snapshots of the evolution of the controlled environment are shown in figure~\ref{fig:burgers_fields}. As can be observed, the agent successfully damps the transported inlet noise following an opposition control strategy.

\begin{table}[h!]
    \footnotesize
    \cprotect\caption{\textbf{Default parameters used for} \verb!burgers-v0!}
    \label{table:burgers_parameters}
    \centering
    \begin{tabular}{rll}
        \toprule
        	\verb!L!			& domain length		& $2$\\
        \verb!u_target!		& target value			& $0.5$\\
	\verb!sigma!		& inlet noise level		& $0.1$\\
	\verb!amp!		& control amplitude		& $10$\\
	\verb!ctrl_pos!		& control position		& $1$\\
        \bottomrule
    \end{tabular}
\end{table}

\begin{figure}
\centering
\begin{tikzpicture}[	trim axis left, trim axis right, font=\scriptsize,
				upper/.style={	name path=upper, smooth, draw=none},
				lower/.style={	name path=lower, smooth, draw=none},]
	\begin{axis}[	xmin=0, xmax=500000, scale=0.75,
				ymin=-6, ymax=0,
				scaled x ticks=false,
				xtick={0, 100000, 200000, 300000, 400000, 500000},
				xticklabels={$0$,$100k$,$200k$,$300k$,$400k$,$500k$},
				legend cell align=left, legend pos=south east,
				legend style={nodes={scale=0.8, transform shape}},
				every tick label/.append style={font=\scriptsize},
				grid=major, xlabel=transitions, ylabel=score]
				
		\legend{no control, \textsc{ppo}, \textsc{td3}}
		
		\addplot[thick, opacity=0.7, dash pattern=on 2pt]	coordinates {(0,-3.5) (500000,-3.5)};
		
		\addplot [upper, forget plot] 				table[x index=0,y index=7] {burgers_ppo.dat };
		\addplot [lower, forget plot] 				table[x index=0,y index=6] {burgers_ppo.dat}; 
		\addplot [fill=blue3, opacity=0.5, forget plot] 	fill between[of=upper and lower];
		\addplot[draw=blue1, thick, smooth] 			table[x index=0,y index=5] {burgers_ppo.dat}; 
		
		\addplot [upper, forget plot] 				table[x index=0,y index=7] {burgers_td3.dat};
		\addplot [lower, forget plot] 				table[x index=0,y index=6] {burgers_td3.dat}; 
		\addplot [fill=green3, opacity=0.5, forget plot] 	fill between[of=upper and lower];
		\addplot[draw=green1, thick, smooth] 		table[x index=0,y index=5] {burgers_td3.dat}; 
			
	\end{axis}
\end{tikzpicture}
\cprotect\caption{\textbf{Score curves for the} \textsc{ppo} \textbf{and the} \textsc{td3} \textbf{algorithms when solving the} \verb!burgers-v0! \textbf{environment.} The dashed line indicates the reward obtained for the uncontrolled case.} 
\label{fig:burgers_score}
\end{figure} 

\clearpage

\subsection{Sloshing}

\subsubsection{Physics}

We consider the resolution of the 1D Saint-Venant equations (or shallow water equations), established in 1871~\cite{saintvenant1871}, which describe a shallow layer of fluid in hydrostatic balance with constant density. This system is consideblack in the context of a mobile water tank of length $L$ subjected to an acceleration $\ddot{y}$, leading to the following equations in the tank referential~\cite{berger2022}:

\begin{equation}
\label{eq:stvenant}
\begin{split}
	\partial_t h 	&= -\partial_x q, \\
	\partial_t q		&= -\partial_x \left( \frac{q^2}{h} + \frac{1}{2} g h^2 \right) - \ddot{y},
\end{split}
\end{equation}

where $h$ is the fluid height, and $q$ is the fluid flow rate. The system (\ref{eq:stvenant}) is completed by the following initial and boundary conditions:

\begin{equation}
\label{eq:stvenant_bc}
\begin{split}
	q(x,0)	= 0 &\text{ and } h(x,0)	= 1, \\
	q(0,t) 	= 0 &\text{ and } \partial_x h(0,t) = 0, \\
	q(L,t) 	= 0 &\text{ and } \partial_x h(L,t) = 0.
\end{split}
\end{equation}

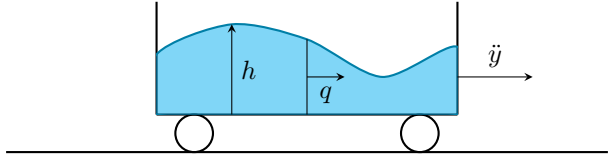
\begin{figure}[h!]
\centering
\begin{tikzpicture}[]

	\draw [thick] (-4,0) -- (4,0);

	\draw [thick] (-2,0.5) -- (2,0.5);
	\draw [thick] (-2,0.5) -- (-2,2);
	\draw [thick] (2,0.5) -- (2,2);

	\draw [thick](-1.5,0.25) circle (0.25cm);
	\draw [thick] (1.5,0.25) circle (0.25cm);

	\begin{scope}
	    	\clip(-2,0.5) rectangle (2,2);
		\draw[draw=bluegray1,fill=cyan,thick,fill opacity=0.5] plot [smooth cycle] coordinates {(-2.0,0.5+0.8) (-1.0,0.5+1.2) (0.0,0.5+1.0) (1.0,0.5+0.5) (2.0,0.5+0.9) (2.0,0.5+0.0) (-2.0,0.5+0.0)};
	\end{scope}

	\draw[-stealth] (-1.0,0.5) -- (-1.0,1.7) node[pos=0.5, anchor=west] {$h$};
	\draw[] (0,0.5) -- (0,1.5);
	\draw[-stealth] (0,1.0) -- (0.5,1.0) node[pos=0.5, anchor=north] {$q$};
	\draw[-stealth] (2.0,1.0) -- (3.0,1.0) node[pos=0.5, anchor=south] {$\ddot{y}$};
	
\end{tikzpicture}
\caption{\textbf{Configuration of the sloshing tank.} The fluid flow is determined by the fluid height $h(x,t)$ and by its mass flow rate $q(x,t)$. The movement of the tank is controlled by its acceleration $\ddot{y}(t)$.} 
\label{fig:sloshing_tank}
\end{figure}

\begin{figure}[h!]
\centering
\begin{subfigure}{0.45\textwidth}
	\centering
	\fbox{\includegraphics[width=\textwidth]{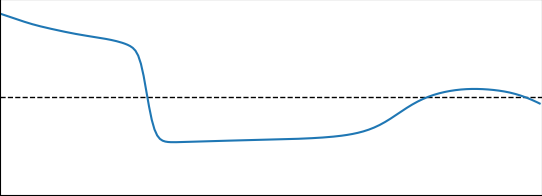}}
    	\caption{Excitation phase}
	\label{fig:sloshing_excitation}
\end{subfigure} \quad
\begin{subfigure}{0.45\textwidth}
	\centering
	\fbox{\includegraphics[width=\textwidth]{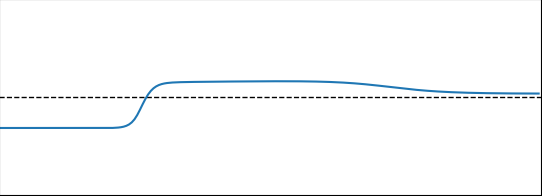}}
    	\caption{Relaxation phase}
	\label{fig:sloshing_free}
\end{subfigure}
\caption{\textbf{Examples of fluid surface during the excitation phase (left) and the relaxation phase (right)}. \textcolor{red}{The horizontal axis represents the $x$ coordinate, while the vertical axis represents the height of the fluid. The horizontal line indicates the height of the fluid at rest ($h=1$).}}
\label{fig:sloshing_examples}
\end{figure}

The situation is summed up in figure~\ref{fig:sloshing_tank}. When laterally excited, the surface of the fluid sloshes back and forth in the tank generating complex patterns at the fluid surface, as shown in figure~\ref{fig:sloshing_examples}. When the excitation stops, a relaxation phase is observed, usually leaving a single wavefront travelling back and forth in the tank until it dissipates entirely. Due to its simplicity, the model (\ref{eq:stvenant}) does not allow wave breaking nor the formation of drops on the sides of the domain.

\subsubsection{Discretization}

The system (\ref{eq:stvenant}) is discretized using a finite volume scheme with Rusanov fluxes~\cite{cordier2007}, which is an improved form of the Lax-Friedrichs flux. The time derivatives for both $h$ and $q$ are discretized using the second-order Adams-Bashforth scheme.

\subsubsection{Environment}

The control of the system (\ref{eq:stvenant}) is performed through the cart acceleration term $\ddot{y}$. The system is first set in motion during $t_\text{exc} = 2$ time units using a sinusoid-based signal:

\begin{equation}
	\ddot{y}_\text{exc} (t) = \frac{1}{2} \left( \cos ( \pi t ) + 3 \cos ( 4 \pi t) \right).
\end{equation}

The resulting fields are stoblack in a file for simplicity, and loaded at the beginning of each episode. By default, the length of the cart is $L = 2.5$, the spatial discretization corresponds to $100$ finite volume cells per unit of length, and the numerical time step is $\Delta t = 0.001$ time units. The actions provided to the environment, which are expected to be in $\left[-1, 1\right]$, are then scaled by an \textit{ad-hoc} non-dimensional amplitude factor $A=5$. The interpolation between successive actions is identical to (\ref{eq:shkadov_actions}), with $\Delta t_\text{int} = 0.01$ time units and $\Delta t_\text{act} = 0.05$ time units. The total episode time is fixed to $10$ time units, corresponding to $200$ actions. The observations provided to the agent are the heights collected on the entire domain. To limit the size of the resulting vector, it is downsampled by a factor 2. Finally, the reward signal is defined as:

\begin{equation}
\label{eq:sloshing_reward}
	r(t) = - \Delta x \norm{h(x,t) - 1}_2 - \alpha \left| \ddot{y} (t) \right|,
\end{equation}

where $\norm{\cdot}_2$ is the $2$-norm and $\alpha = 0.0005$. The $\Delta x$ factor allows to obtain comparable reward values for variable discretization levels.

\subsubsection{Results}

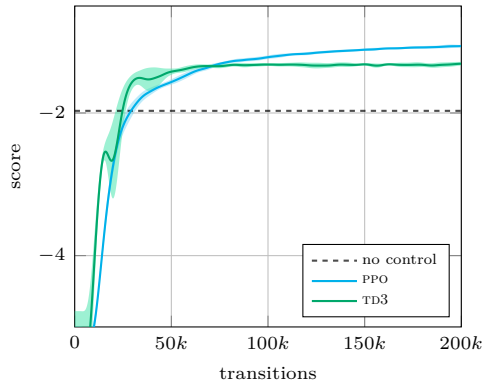
\begin{figure}
\centering
\begin{tikzpicture}[	trim axis left, trim axis right, font=\scriptsize,
				upper/.style={	name path=upper, smooth, draw=none},
				lower/.style={	name path=lower, smooth, draw=none},]
	\begin{axis}[	xmin=0, xmax=200000, scale=0.75,
				ymin=-5, ymax=-0.5,
				scaled x ticks=false,
				xtick={0, 50000, 100000, 150000, 200000},
				xticklabels={$0$,$50k$,$100k$,$150k$,$200k$},
				legend cell align=left, legend pos=south east,
				legend style={nodes={scale=0.8, transform shape}},
				every tick label/.append style={font=\scriptsize},
				grid=major, xlabel=transitions, ylabel=score]
				
		\legend{no control, \textsc{ppo}, \textsc{td3}}
		
		\addplot[thick, opacity=0.7, dash pattern=on 2pt]	coordinates {(0,-1.97) (200000,-1.97)};
		
		\addplot [upper, forget plot] 				table[x index=0,y index=7] {sloshing_ppo.dat};
		\addplot [lower, forget plot] 				table[x index=0,y index=6] {sloshing_ppo.dat}; 
		\addplot [fill=blue3, opacity=0.5, forget plot] 	fill between[of=upper and lower];
		\addplot[draw=blue1, thick, smooth] 			table[x index=0,y index=5] {sloshing_ppo.dat};
		
		\addplot [upper, forget plot] 				table[x index=0,y index=7] {sloshing_td3.dat};
		\addplot [lower, forget plot] 				table[x index=0,y index=6] {sloshing_td3.dat}; 
		\addplot [fill=green3, opacity=0.5, forget plot] 	fill between[of=upper and lower];
		\addplot[draw=green1, thick, smooth] 		table[x index=0,y index=5] {sloshing_td3.dat}; 
			
	\end{axis}
\end{tikzpicture}
\cprotect\caption{\textbf{Score curves for the} \verb!sloshing-v0! \textbf{environment using the} \textsc{ppo} \textbf{and the} \textsc{td3} \textbf{algorithms.} The dashed line indicates the reward obtained for the uncontrolled environment.} 
\label{fig:sloshing_score}
\end{figure}

\begin{figure}
\centering

\begin{subfigure}[t]{\textwidth}
	\centering
	\fbox{\includegraphics[width=.4\textwidth]{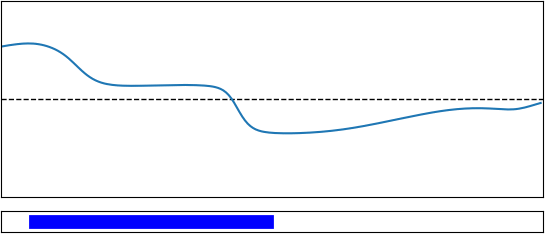}} \hspace{.5cm} \fbox{\includegraphics[width=.4\textwidth]{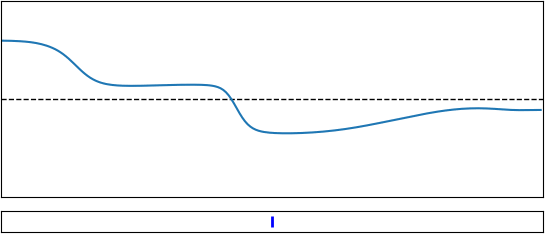}}
    	\caption{$t=0$}
	\label{fig:sloshing_field_0}
\end{subfigure}

\medskip

\begin{subfigure}[t]{\textwidth}
	\centering
	\fbox{\includegraphics[width=.4\textwidth]{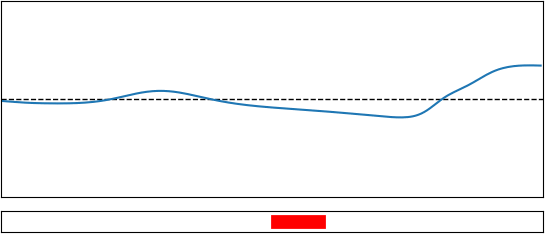}} \hspace{.5cm} \fbox{\includegraphics[width=.4\textwidth]{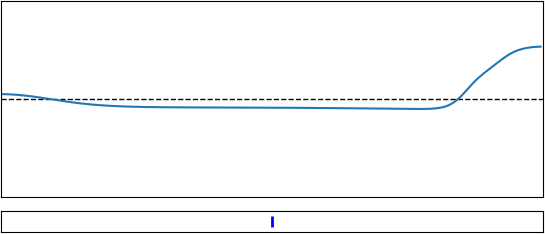}}
    	\caption{$t=0.5$}
	\label{fig:sloshing_field_10}
\end{subfigure} 

\medskip

\begin{subfigure}[t]{\textwidth}
	\centering
	\fbox{\includegraphics[width=.4\textwidth]{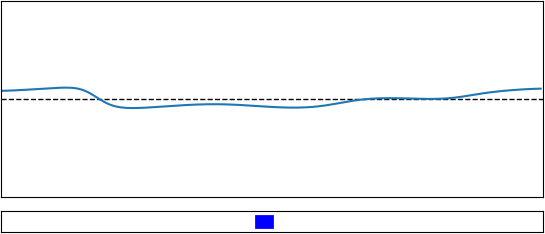}} \hspace{.5cm} \fbox{\includegraphics[width=.4\textwidth]{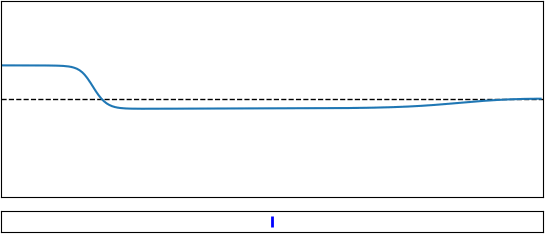}}
    	\caption{$t=1.5$}
	\label{fig:sloshing_field_30}
\end{subfigure} 

\medskip

\begin{subfigure}[t]{\textwidth}
	\centering
	\fbox{\includegraphics[width=.4\textwidth]{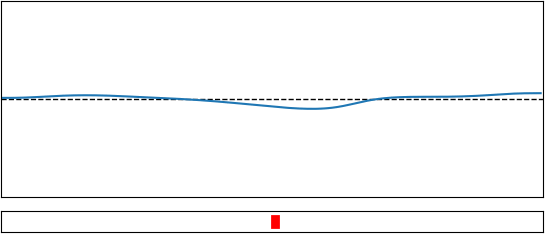}} \hspace{.5cm} \fbox{\includegraphics[width=.4\textwidth]{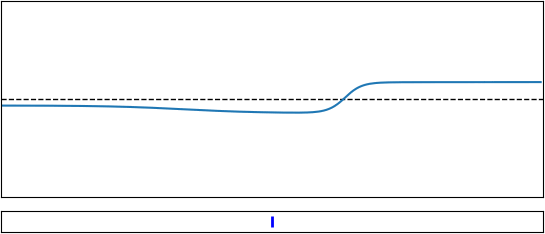}}
    	\caption{$t=2.5$}
	\label{fig:sloshing_field_50}
\end{subfigure} 

\medskip

\begin{subfigure}[t]{\textwidth}
	\centering
	\fbox{\includegraphics[width=.4\textwidth]{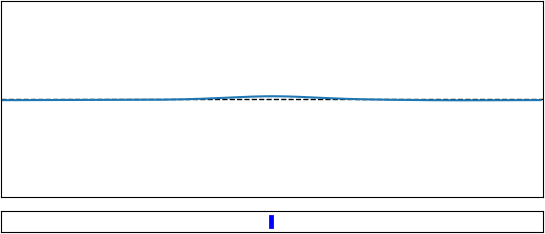}} \hspace{.5cm} \fbox{\includegraphics[width=.4\textwidth]{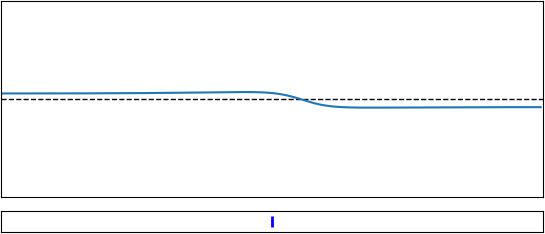}}
    	\caption{$t=5$}
	\label{fig:sloshing_field_100}
\end{subfigure} 

\medskip

\begin{subfigure}[t]{\textwidth}
	\centering
	\fbox{\includegraphics[width=.4\textwidth]{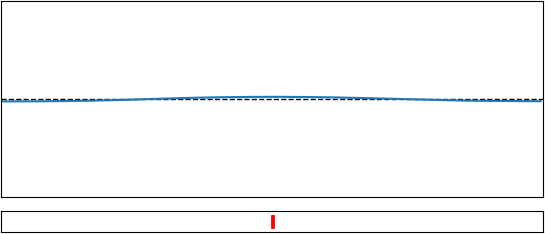}} \hspace{.5cm} \fbox{\includegraphics[width=.4\textwidth]{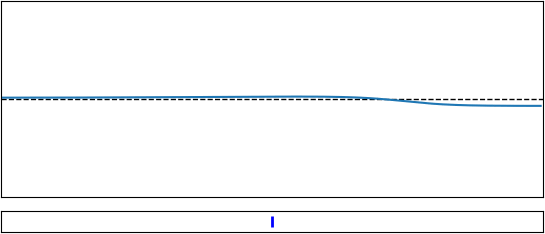}}
    	\caption{$t=10$}
	\label{fig:sloshing_field_10}
\end{subfigure} 


\caption{\textbf{Evolution of the fluid surface with (left) and without (right) agent control.} The control amplitude and direction is represented using the rectangle at the bottom (red means positive, blue means negative). \textcolor{red}{The axes are the same as in figure \ref{fig:sloshing_examples}.}}
\label{fig:sloshing_fields}
\end{figure}

The environment as described in the previous section is referblack to as \verb!sloshing-v0!, and its default parameters are provided in table~\ref{table:sloshing_parameters}. \textcolor{black}{In this section, we note that the critic learning rate of the \textsc{ppo} agent is blackuced to $\lambda_c = \num{1e-3}$ compablack to the default hyperparameters of table~\ref{table:default_ppo_parameters}.} For the training, we set $n_\text{rollout} = 2000$, $n_\text{batch} = 2$, $n_\text{epoch} = 16$ and $n_\text{max} = 200k$. The score curves are presented in figure~\ref{fig:sloshing_score}, while the time evolutions of the controlled versus uncontrolled fluid level are shown in figure~\ref{fig:sloshing_fields}. As can be observed, the agent manages to roughly cut the uncontrolled reward in half, by suppressing the back and forth wavefront using large actuations in the early stages of control, after what the control amplitude drops significantly.

\begin{table}[h!]
    \footnotesize
    \cprotect\caption{\textbf{Default parameters used for} \verb!sloshing-v0!}
    \label{table:sloshing_parameters}
    \centering
    \begin{tabular}{rll}
        \toprule
        \verb!L!			& length of the tank						& $2.5$\\
	\verb!amp!		& amplitude of the control					& $5$\\
	\verb!alpha!		& control penalization					& $0.0005$\\
	\verb!g!			& gravity acceleration					& $9.81$\\
        \bottomrule
    \end{tabular}
\end{table}

\clearpage

\subsection{Vortex}

\subsubsection{Physics}

We consider the resolution of a dynamical system modeling the nonlinear vortex-induced vibrations of a rigid circular cylinder. The flow motion is governed by the incompressible Navier--Stokes equations, whereas the cylinder motion is a simple translation governed by a linear mass-damper-spring equation affected by the fluid loading. This is modeled by coupled amplitude equations derived in~\cite{meliga2011a,meliga2011b} after dominant balance arguments:

\begin{equation}
\label{eq:vortex}
\begin{split}
\partial_t A 	&= \lambda \left( \frac{1}{\re_*} - \frac{1}{\re} \right) A - \mu A \lvert A \rvert^2 + \alpha Y \\
\partial_t Y 	&= \left( -\omega_* \gamma + \iu \left( \omega_s - \omega_* \right) \right) Y + \frac{\beta}{\omega_* m} A
\end{split}
\end{equation}

where $A$ and $Y$ are unknown slow time-varying, complex amplitudes modeling respectively the flow disturbances and the cylinder center of mass, $\re = 50$ is the Reynolds number, $\re_* = 46.6$ is the threshold of instability of the steady cylinder, $\omega_* = 0.74$ is the frequency of the marginally stable eigenmode classically computed from the flow past a fixed cylinder~\cite{barkley2006}, $\omega = 1.1$ is the dimensionless natural frequency of the cylinder in vacuum, $\gamma = 0.023$ is the structural damping coefficient, and $m = 10$ is the ratio of the solid to the fluid densities. The coefficients $\lambda$, $\mu$, $\alpha$, $\beta$ in (\ref{eq:vortex}) are analytically computable from an asymptotic analysis of the coupled flow-cylinder system, their numerical value being taken from~\cite{meliga2011a} as:

\begin{equation}
\label{eq:values}
\lambda = 9.153 + 3.239 \iu, \quad \mu = 308.9 - 1025 \iu, \quad \alpha = 0.03492 + 0.01472 \iu, \quad \beta=1.
\end{equation}

The ability of the model to reproduce the physics of vortex-induced vibrations has been assessed from the study of the nonlinear limit cycles, \textit{i.e.} the periodic, synchronized orbits reached by the system at large time whose analytical expressions are reported in~\cite{meliga2011a}. Of particular importance is the simultaneous existence of multiple stable cycles (either a single limit cycle, or three over specific ranges of frequencies), that is shown to trigger a complex hysteretic behavior in the lock-in regime. Since all limit cycle solutions are periodic, the mean mechanical energy averaged over a period is zero, and the mean work received from the fluctuating lift force is entirely dissipated by structural damping. As discussed in~\cite{meliga2011a}, it follows that the leading order mean dissipated energy is a simple quadratic function of the displacement amplitude, meaning that only the upper limit cycle (the one limit cycle yielding the largest displacement amplitude) is of practical interest for the energy extraction problem, that is, in cases where one seeks to leverage such vortex-induced vibrations to generate electrical energy, for instance by having the oscillation of the cylinder displace periodically a magnet within a coil.

In essence, it can be inferblack that there must exist an optimal structural parameter setting for which the dissipated energy is maximum: on the one hand, the flow-cylinder system must be synchronized for the cylinder displacement amplitude to be large. On the other hand, the energy tends to zero in the limit $\gamma\gg1$ where the work received from the lift force is limited by the low amplitude of the displacement, and in the limit $\gamma\ll1$ where the displacement is self-limited. As evidenced in~\cite{meliga2011b}, the problem shown is that the optimum lies at the edge of a discontinuity, corresponding to parameter settings where the system undergoes a transition from a hysteretic to a non-hysteretic regime. This has important consequences for the application, since small inaccuracies in the structural parameters of small external flow disturbances may tip the system outside the hysteresis zone and lead to convergence to cycles of lower energy, resulting in a dramatic drop of the harnessed energy.

\subsubsection{Discretization}

The system (\ref{eq:vortex}) is integrated in time using a five-stage, fourth-order low-storage Runge Kutta scheme from Carpenter \textit{et al.}~\cite{carpenter1994}, using a constant time-step $\Delta t=0.1$.

\subsubsection{Environment}

The proposed environment is re-implemented based on the original work of~\cite{meliga2011b}, where the goal is to maximize the cylinder displacement and bypass the existence of low energy cycles. This is achieved adding a proportional feedback control in the structure equation, assuming that the state of the system is accessed through measurements of the flow disturbances position, and that an actuator applies at the surface of the cylinder a control velocity:

\begin{equation}
\label{eq:vortex}
\begin{split}
\partial_t A &= \lambda \left( \frac{1}{\re_*} - \frac{1}{\re} \right) A - \mu A \lvert A \rvert^2 + \alpha Y + k e^{ \iu \phi } A \\
\partial_t Y &= \left( -\omega_* \gamma + \iu \left( \omega_s - \omega_* \right) \right) Y + \frac{\beta}{\omega_* m} A
\end{split}
\end{equation}

with $k$ the gain and $\phi$ the phase shift between the measure and the action. The actions provided to the environment, which are expected to be in $\left[-1, 1\right]$, are rescaled to $\left[0, 0.3\right]$ for the module, and $\left[-\pi, \pi \right]$ for the phase. The action time-step is set to $\Delta t_\text{act} = 0.5$ time units, a full episode lasting $400$ time units, corresponding to $800$ actions.  The observations are the complex variables $A$ and $Y$ as well as their time-derivatives, leading to an observation vector of size $8$. The reward at each time-step is computed as:

\begin{equation}
r(t) = 8 \omega_s \gamma \left( \frac{\text{d}}{\text{d}t}  \real \left[ Y e^{\iu \omega_* t} \right] \right)^2 - 2 w \real \left[ k e^{\iu \phi} A e^{\iu \omega_* t} \right]^2,
\end{equation}

where the leftmost term is the mean dissipated energy and is thus associated to performance, the rightmost term estimates the mean kinetic energy expended by the actuator over a limit cycle period and is thus associated to cost, and $w$ is a weighting coefficient set empirically to $50$ (a value found to be large enough for cost considerations to impact the optimization procedure, but not so large as to dominate the reward signal, in which case actuating is meaningless). Each episode begins using the same initial condition $(A_0, Y_0) = \left( -0.00385 - 0.00378 \iu, 0.00118 - 0.00131 \iu \right)$ that, in the absence of control, leads convergence to a low limit cycle, for which the score is equal to $0.00607$.

\subsubsection{Results}

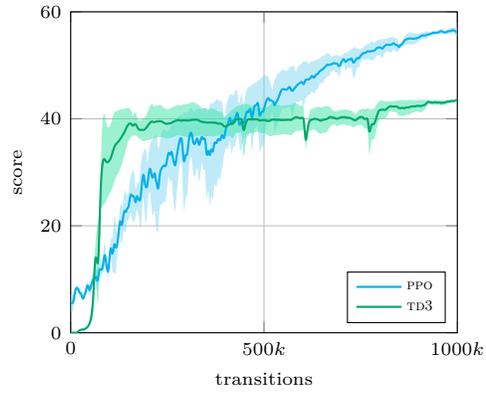
\begin{figure}
\centering
\begin{tikzpicture}[	trim axis left, trim axis right, font=\scriptsize,
				upper/.style={	name path=upper, smooth, draw=none},
				lower/.style={	name path=lower, smooth, draw=none},]
	\begin{axis}[	xmin=0, xmax=1000000, scale=0.75,
				ymin=0, ymax=60,
				scaled x ticks=false,
				xtick={0, 500000, 1000000},
				xticklabels={$0$,$500k$,$1000k$},
				legend cell align=left, legend pos=south east,
				legend style={nodes={scale=0.8, transform shape}},
				every tick label/.append style={font=\scriptsize},
				grid=major, xlabel=transitions, ylabel=score]
				
		\legend{\textsc{ppo}, \textsc{td3}}
		
		
		\addplot [upper, forget plot] 				table[x index=0,y index=7] {vortex_ppo.dat};
		\addplot [lower, forget plot] 				table[x index=0,y index=6] {vortex_ppo.dat}; 
		\addplot [fill=blue3, opacity=0.5, forget plot] 	fill between[of=upper and lower];
		\addplot[draw=blue1, thick, smooth] 			table[x index=0,y index=5] {vortex_ppo.dat};
		
		\addplot [upper, forget plot] 				table[x index=0,y index=7] {vortex_td3.dat};
		\addplot [lower, forget plot] 				table[x index=0,y index=6] {vortex_td3.dat}; 
		\addplot [fill=green3, opacity=0.5, forget plot] 	fill between[of=upper and lower];
		\addplot[draw=green1, thick, smooth] 		table[x index=0,y index=5] {vortex_td3.dat}; 
			
	\end{axis}
\end{tikzpicture}
\cprotect\caption{\textbf{Score curves for the} \verb!vortex-v0! \textbf{environment using the} \textsc{ppo} \textbf{and the} \textsc{td3} \textbf{algorithms.} The uncontrolled environment obtains a score of $0.00607$ when locking in on the low limit cycle.} 
\label{fig:vortex_score}
\end{figure}

\begin{figure}
\centering
\pgfdeclarelayer{background}
\pgfsetlayers{background,main}

\begin{subfigure}[t]{.3\textwidth}
	\centering
	\fbox{\includegraphics[width=\textwidth]{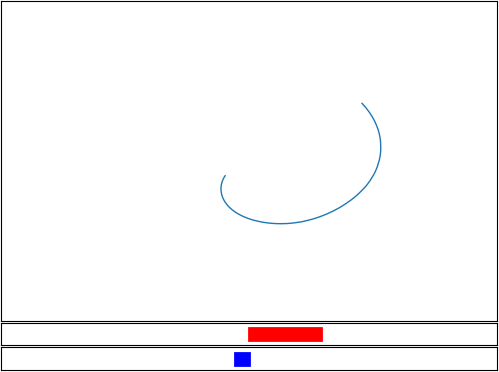}}
    	\caption{$t=25$}
	\label{fig:vortex_field_25}
\end{subfigure} \quad
\begin{subfigure}[t]{.3\textwidth}
	\centering
	\fbox{\includegraphics[width=\textwidth]{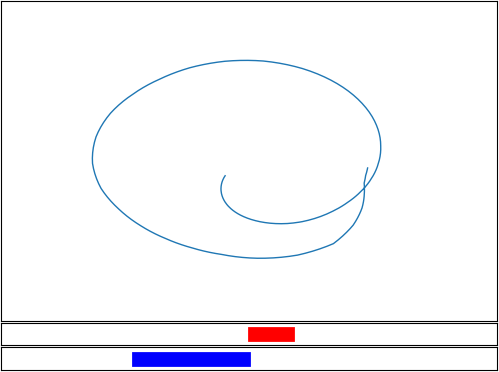}}
    	\caption{$t=50$}
	\label{fig:vortex_field_50}
\end{subfigure} \quad
\begin{subfigure}[t]{.3\textwidth}
	\centering
	\fbox{\includegraphics[width=\textwidth]{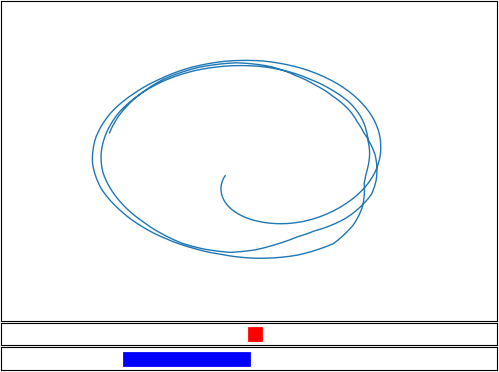}}
    	\caption{$t=100$}
	\label{fig:vortex_field_100}
\end{subfigure}

\medskip

\begin{subfigure}[t]{.3\textwidth}
	\centering
	\fbox{\includegraphics[width=\textwidth]{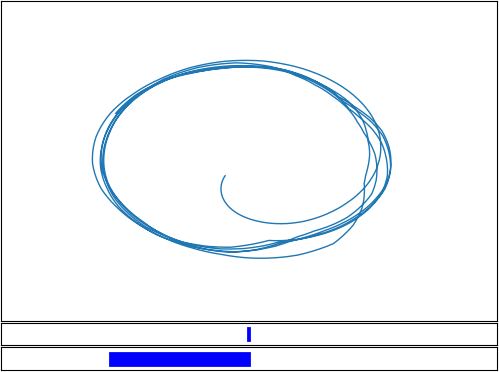}}
    	\caption{$t=200$}
	\label{fig:vortex_field_200}
\end{subfigure} \quad
\begin{subfigure}[t]{.3\textwidth}
	\centering
	\fbox{\includegraphics[width=\textwidth]{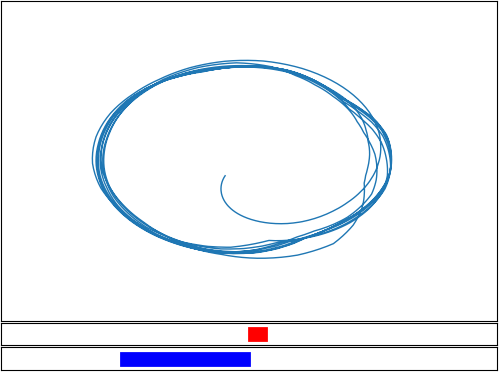}}
    	\caption{$t=300$}
	\label{fig:vortex_field_300}
\end{subfigure} \quad
\begin{subfigure}[t]{.3\textwidth}
	\centering
	\fbox{\includegraphics[width=\textwidth]{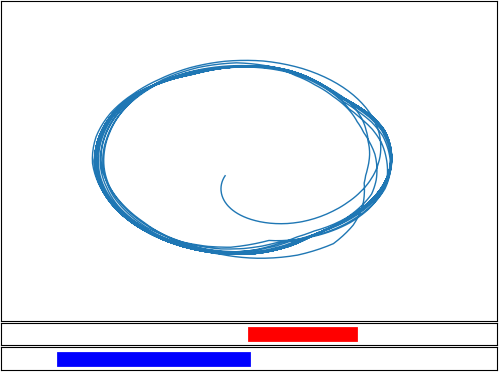}}
    	\caption{$t=800$}
	\label{fig:vortex_field_800}
\end{subfigure}

\caption{\textbf{Evolution of the controlled vortex system} using the \textsc{ppo} algorithm. The instantaneous control values are indicated at the bottom by the colored bars.}
\label{fig:vortex_fields}
\end{figure}

The environment as described in the previous section is referblack to as \verb!vortex-v0!, and its default parameters are provided in table~\ref{table:vortex_parameters}. \textcolor{black}{In this section, we note that the critic learning rate of the \textsc{ppo} agent is blackuced to $\lambda_c = \num{1e-3}$ compablack to the default hyperparameters of table~\ref{table:default_ppo_parameters}.} For the training, we set $n_\text{rollout} = 4000$, $n_\text{batch} = 2$, $n_\text{epoch} = 32$ and $n_\text{max} = 1000k$. The score curves are presented in figure~\ref{fig:vortex_score}, while the time evolutions of the controlled versus uncontrolled fluid level are shown in figure~\ref{fig:vortex_fields}. As can be observed, the agent manages to lock in a high limit cycle, with a final score more than 3 orders of magnitude larger than that of the uncontrolled low limit cycle, and twice as large as that of the uncontrolled high limit cycle (whose score computed using  $A_0 = Y_0 = 0.05(1+\iu)$ is $19.25$).

\begin{table}[h!]
    \footnotesize
    \cprotect\caption{\textbf{Default parameters used for} \verb!vortex-v0!}
    \label{table:vortex_parameters}
    \centering
    \begin{tabular}{rll}
        \toprule
        \verb!re!			& Reynolds number						& $50$\\
	\verb!w!			& control penalization					& $50$\\
        \bottomrule
    \end{tabular}
\end{table}

\clearpage

\section{Conclusion and availability}

\textcolor{black}{In the present work seven fluid dynamics environments are proposed, each of them being representative of a specific physical phenomenon. We observed that the performance ranking usually observed in standard benchmarks such as \textsc{gym} and \textsc{mujoco} (\textit{i.e.} that off-policy methods usually significantly outperform on-policy approaches) is not necessarily valid in the context of flow control environments.}

\textcolor{black}{The environments are implemented in a modular way, following the structure of the \textsc{gym} library to allow for a seamless integration with existing code. Optimizations have been run on each of the cases and the results have been commented in each of the respective sections. The sources of the present work are made open-source on the following repository: \url{https://github.com/jviquerat/beacon}.}

\textcolor{black}{While the present version of the library presents a modest variety of phenomena and control types, its purpose is to grow with new cases proposed by the community, within the constraints detailed in section \ref{section:lib}. To this end, issues and pull requests are accepted on the library repository. With the present work, we hope to provide a solid foundation for the development of a community-driven library of fluid dynamics environments, and to foster the development of new control strategies for fluid dynamics.}

\section*{Funding}

Funded/co-funded by the European Union (ERC, CURE, 101045042). Views and opinions expressed are however those of the authors only and do not necessarily reflect those of the European Union or the European Research Council. Neither the European Union nor the granting authority can be held responsible for them.

\appendix

\section{Implementations benchmarks}
\label{section:benchmark}

\textcolor{black}{This section is dedicated to the evaluation of the in-house \textsc{ppo} and \textsc{td3} algorithms on standard benchmarks. In this regard, we selected two cases from the \textsc{gym} library, and two cases from the \textsc{mujoco} library. Results are presented in figure \ref{fig:ppo_perf}. The implementations used to solve the \textsc{beacon} benchmark cases compare favourably with reference implementations.}

\input{ppo_perf}

\bibliographystyle{unsrt}
\bibliography{bib}

\end{document}